%% file: plain.tex
\documentclass[]{article}
\usepackage[backend=biber]{biblatex}

\usepackage[paper=a4paper,left=25mm,right=25mm,top=25mm,bottom=25mm]{geometry}

\usepackage[english]{babel}
\usepackage[english=american]{csquotes} 

\usepackage{tikz}
\usepackage{multirow}
\usepackage{pstricks}    
\usepackage{graphicx}    
\usepackage{epstopdf}

\usepackage{stfloats}
\usepackage{url}

\usepackage{subcaption}

\usepackage{fancyhdr}

\usepackage{framed}

\usepackage{authblk}

\usepackage{ifthen}

\begin{document}
\title{Spreadsheet Guardian: An Approach to Protecting Semantic Correctness throughout the Evolution of Spreadsheets}

\author[1]{Daniel Kulesz}
\author[1]{Verena K\"afer}
\author[1]{Stefan Wagner}
\affil[1]{Software Engineering Group, Institute of Software Technology, University of Stuttgart, Germany}

\date{}

\pagestyle{fancy}

\fancyhead{}
\fancyhead[L]{\ifthenelse{\isodd{\value{page}}}{Spreadsheet Guardian: Protecting semantic correctness throughout the evolution of spreadsheets}{Daniel Kulesz, Verena K\"afer and Stefan Wagner}}

\maketitle
\begin{abstract}
\input{abstract}

\end{abstract}



\input{mainpaper-preprint.tex}

\newpage


\printbibliography

\end{document}

%% file: abstract.tex
Spreadsheets are powerful tools which play a business-critical role in many organizations. However, many bad decisions taken due to faulty spreadsheets show that these tools need serious quality assurance. Furthermore, while collaboration on spreadsheets for maintenance tasks is common, there has been almost no support for ensuring that the spreadsheets remain correct during this process.

We have developed an approach named Spreadsheet Guardian which separates the specification of spreadsheet test rules from their execution. By automatically executing user-defined test rules, our approach is able to detect semantic faults. It also protects all collaborating spreadsheet users from introducing faults during maintenance, even if only few end-users specify test rules. To evaluate Spreadsheet Guardian, we implemented a representative testing technique as an add-in for Microsoft Excel.

We evaluated the testing technique in two empirical evaluations with 29 end-users and 42 computer science students. The results indicate that the technique is easy to learn and to apply. Furthermore, after finishing maintenance, participants with spreadsheets \enquote{protected} by the technique are more realistic about the correctness of their spreadsheets than participants who employ only \enquote{classic}, non-interactive test rules based on static analysis techniques. Hence, we believe Spreadsheet Guardian can be of use for business-critical spreadsheets.

%% file: mainpaper-preprint.tex
\newcommand{\confnote}{\footnote{We provide an example video for the reviewers of our submission. It is available from the following URL: https://www.hidrive.strato.com/lnk/wILCppDz with password "spreadsheet-guardian". We also provide an archive with all screen recordings from E3 (11GB) for the reviewers from the following URL: https://www.hidrive.strato.com/lnk/zfLCJGMt}}
\renewcommand{\confnote}{}

\section{Introduction}

Although spreadsheets have existed for more than 35 years, their popularity remains unbroken: it is assumed that there is a population of millions of spreadsheet users and that billions of spreadsheets exist \cite{scaffidi2005estimating}. Since the 1980s, spreadsheets have played a critical role in most businesses -- they are used for accounting \cite{marriott2004using}, data analysis \cite{ragsdale2004spreadsheet}, as decision support systems \cite{weitze2014spreadsheet} and for many other purposes \cite{grossman2007lessons}.

\subsection{Motivation}

Using untested spreadsheets can be risky. Over the last years, dozens of cases were disclosed where faulty spreadsheets caused severe financial and reputational damage \cite{horror}, with the recent Rogoff-Reinhart study being one of the most drastic examples \cite{herndon2014does, reinhart2010growth}. 

Despite these issues and some conceptional shortcomings of spreadsheets, no other technology has managed to supplant spreadsheets on a broader scale. For these reasons -- and also due to the number of existing spreadsheets -- numerous scientists have investigated the creation, detection and prevention of faults in spreadsheets \cite{jannach2014avoiding, panko1998we}. Yet, we see three areas that require more attention: overconfidence, collaboration and tool support.

\emph{Overconfidence}: Governmental institutions have already taken action to fight faulty spreadsheets: several recent laws like the Sarbanes Oxley Act 404, Basel III or Solvency II demand proof from organizations that all artifacts which contribute to financial calculations have been thoroughly inspected -- which also includes spreadsheets used in the process.
However, Spreadsheet inspection approaches are not very helpful if spreadsheet users lack the awareness of why inspections are necessary. As a consequence, inspections are not executed to the right extent, not at the right point in time or not at all. Creating such an awareness is a non-trivial task that can be difficult and tedious -- especially considering the issue of overconfidence \cite{panko2003reducing, panko2007two}. We argue that reducing overconfidence is even more important than increasing correctness as many faulty spreadsheets which caused severe damage did not undergo any inspections at all.

\emph{Collaboration}: When it comes to maintenance of spreadsheets, a highly relevant but often ignored factor is collaboration among end-users. In 2005 and 2006, researchers at Dartmouth College issued a comprehensive study about spreadsheet work habits \cite{baker2006survey}. A total of 1,597 MBA graduates from 7 samples responded to an online questionnaire. For the majority of them, spreadsheets play a very important (33.6\%) or critical (49\%) role in their work. According to the study, most spreadsheet users (81.1\%) work on their spreadsheets alone but when creating new spreadsheets, they often (62.1\%) use existing spreadsheets as templates. Also, it is rather uncommon that spreadsheets are used only by their authors: they are shared with one or two (42\%) or more (30.9\%) other persons. When sharing spreadsheets, users typically share the whole file (67.6\%).  With these numbers in mind, it is alarming that 88.1\% of the participants of this study stated that they do not invest any time in 
documenting their spreadsheets! An older study by Nardi and Miller shows that collaboration in spreadsheet environments was already common in the 1990s \cite{nardi1991twinkling}.

\emph{Tool support}: The fact that collaboration among end-users is very common increases the need for tool support around this activity. Even when an existing spreadsheet is just copied, it is quite likely that the intentions of the original spreadsheet are unknown to the copying user and therefore it is easily possible that the new spreadsheet will be used unchanged for the wrong purpose. Also, when changing or improving an existing spreadsheet, it is possible to damage it, for example by expanding it in a wrong way or by damaging existing formulas. One big problem with this is the fact that current spreadsheet execution environments (such as Microsoft Excel) can check if a formula is syntactically correct, but they cannot know if the logic behind the formula makes any sense.

Microsoft Excel provides a feature that shows a marker if a formula is different from the ones in the surrounding cells, but this can only help in very uniformly structured spreadsheets and cannot be applied for more unstructured ones. Without the possibility for the user to manually enter tests for formulas, there is currently no way how a spreadsheet execution environment can check if a formula is correct. Another feature provided by Microsoft Excel, LibreOffice Calc and Google Sheets is the data validation mechanism that can alert users if they enter data that does not meet user-specifiable conditions. The feature can be applied to the result of formulas as well and can be configured more strictly to even reject invalid data instead of only giving warnings. The major drawback of this feature is that it only allows to enter one condition to be checked. Also, the implementation of this feature in Microsoft Excel 2013 and LibreOffice Calc 5.2 triggers the checks only on the cell the user changed last: If the 
user once dismissed a warning he will not be alerted again until he or she changes the cell. Also, if the mechanism is applied to a formula and the user changes a cell the formula depends on, the user is not alerted either.

\subsection{Problem Statement}

Maintenance activities in spreadsheets typically are not clearly separated from \enquote{normal} usage activities. Furthermore, because spreadsheet users are often overconfident about the correctness of their spreadsheets \cite{panko2003reducing}, they do not undertake thorough quality assurance activities during and after maintenance activities. The fact that spreadsheet users often collaborate and share their undocumented and untested spreadsheets with co-workers further increases the risk of running into problems. Additionally, current spreadsheet execution environments (and other programming environments) can only check the syntactic correctness of formulas but not their semantic correctness.

\subsection{Research Objective}

Since spreadsheets have many similarities with traditional programs, it seems worthwhile considering to port proven insights from software engineering to the world of spreadsheets. However, spreadsheets require dealing with collaborative maintenance activities in often uncontrolled arbitrary environments. Therefore, the main objective of this work was to find a concept which can increase semantic correctness in such settings while being unintrusive and keeping the need for changing users' habits as low as possible.

\subsection{Contributions}

This work presents an approach called \textit{Spreadsheet Guardian} which ports ideas from continuous software engineering to the world of spreadsheets. The central idea of Spreadsheet Guardian is that end-users specify tests which, once specified, are continuously and automatically executed in the background. With the tests in place, all current and future users of a \enquote{protected} spreadsheet are warned if they change the spreadsheet in a way that violates its specification (implicitly expressed by the test rules). Yet, the specification and execution of the tests are separated. Spreadsheet Guardian is conceptually open to support any technique that allows the specification of test rules that can be applied at runtime and is not limited to unit tests.

In previous work \cite{kulesz2014integrating}, we have presented a tool named \textit{Spreadsheet Inspection Framework} (SIF). SIF is tightly integrated into Microsoft Excel and allows end-users to create unit-test-like test rules. Once specified, SIF executes these test rules continuosly in the background and reports findings to its users. Thus, SIF is the first proof-of-concept implementation of the Spreadsheet Guardian concept.

While our previous work focused on SIF itself, we have not presented Spreadsheet Guardian as a conceptual approach yet. This work closes this gap by providing three novel contributions: (i) we propose Spreadsheet Guardian as an approach for supporting spreadsheet maintenance activities, (ii) we provide a theory that explains why we think Spreadsheet Guardian is effective and (iii) we present results from two empirical studies where SIF was used as a vehicle to evaluate Spreadsheet Guardian's propositions in practice.

\section{Basics and Terminology}
\label{sec:background}

To make it easier to understand Spreadsheet Guardian and related works, we will introduce some basics and specify our understanding of certain terms in this context.

\subsection{End-User Development}

The interdisciplinary field of End-User Development (EUD) unites End-User Computing \cite{nardi1993small}, End-User Programming \cite{goodell1997end} and End-User Software Engineering \cite{ko2011state}. EUD deals with technology that empowers domain experts to craft and use their own programs instead of using traditional software built for them by somebody else. The common understanding is that domain experts typically do not have a background in computer science or software engineering and are not willing to learn or do programming unless they are convinced of the benefits of doing it for solving a particular task in their domain.

In principle, EUD started in the 1980s with the technological focus on desktop computers in workplaces. Over the years, the focus shifted to personal computers and is currently targeting mobile devices and ubiquitous computing \cite{maceli2017tools}. Even though many researchers nowadays focus on smart homes and IoT, spreadsheets are the \enquote{classic} example of programs crafted by end-users. The underlying principle of having a tabular grid with cells containing data or formulas is still the same as in 1979's VisiCalc \cite{bricklin1999visicalc}, even though VisiCalc itself was superseded by competing products just a few years later \cite{grad2007creation}. Although spreadsheets require spreadsheet execution environments (such as Microsoft Excel) to be created and executed, they can be compared with other interpreted, \enquote{near-standalone} programs such as shell scripts. As technology has moved on, nowadays it is possible to run spreadsheets in the cloud or on mobile devices. This advance makes the 
socio-technical aspects related to collaboration on spreadsheets more important than ever.

\subsection{Spreadsheet Users and Activities}

In contrast to traditional programs, the spreadsheet paradigm does not distinguish between developers and end-users. Therefore, we also use the general term \enquote{spreadsheet user}, but differentiate between various activities of spreadsheet users:

\begin{description}
\item [Development activities] refers to creating, modifying and deleting cells that contain formulas.
\item [Filling activities] refers to filling in data into cells which do not contain formulas.
\item [Viewing activities] refers to reading and interpreting spreadsheets without changing cell contents (neither data nor formulas) for the purpose of getting information.
\item [Inspection activities] refers to reading and interpreting spreadsheets without changing cell contents (neither data nor formulas) for the purpose of understanding the spreadsheet or reviewing it, i.e. checking its correctness.
\end{description}

Depending on the activity, spreadsheet users need different levels of support as Hermans et al. have concluded after surveying 47 employees in an asset management company \cite{hermans2011supporting}.

\subsection{Spreadsheet Errors and Anomalies}

In the spreadsheet literature, the ambiguous term \enquote{spreadsheet error} is used often and several taxonomies for classifying spreadsheet errors have been proposed. Powell et al. \cite{powell2008critical} mention classification possibilities based on:

\begin{itemize}

\item Cause (typing error, copy-paste error,  \ldots)
\item Effect (wrong result, impeded maintainability, \ldots)
\item Type (wrong formula, redundant input fields, \ldots)

\item Development stage (conception, implementation, use)
\end{itemize}

\begin{figure}[htbp]
 \centering
 \includegraphics[width=3.5in]{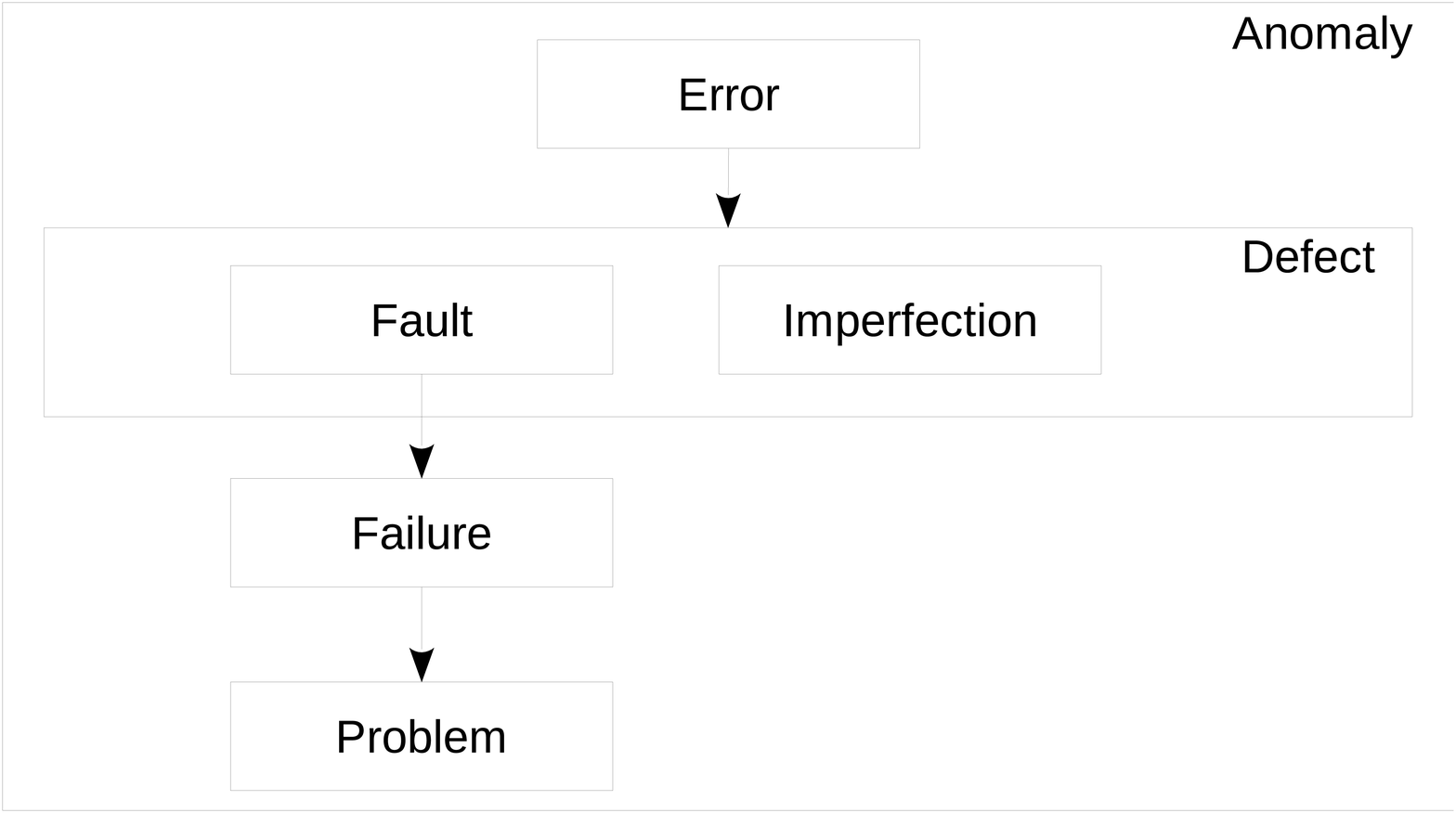}
 \caption{Taxonomy of spreadsheet anomalies}
 \label{fig:taxonomy}
\end{figure}

These possibilities make it clear that the understanding of the term \enquote{spreadsheet error} can be very ambiguous. To avoid this issue, we adopt the notion of an \enquote{anomaly} (that is defined in IEEE Std. 1044:2009 for traditional software) for spreadsheets (see Figure \ref{fig:taxonomy}). In the following, we will explain the various types of anomalies using the following example: Assuming we have a spreadsheet with the following formula in cell C1:

\(=B2+B3+B4+B4+B5+B6+B7+B8\)

\begin{description}
\item [Error] refers to human errors. In the above example, a person could have clicked twice on the cell B4 unintentionally when constructing the formula.
\item [Defect] refers to any undesired data, formula or formatting in a spreadsheet -- such as the formula given in the example.
\item [Fault] refers to a defect that can have a quantitative impact, i.e. on the correctness of the spreadsheet. This would be the case if, in the example above, it was not intended for B4 to be included twice in the result.
\item [Imperfection] refers to a defect that can have a purely qualitative (not quantitative) impact, i.e. on the usability or the maintainability of the spreadsheet. In the example above, this would be the case if summing up B4 twice was actually intended. Also, one could argue that using the SUM-function would be more appropriate here, as well as placing the result of the sum in B9 instead of C1.
\item [Failure] refers to a spreadsheet which is not correct, i.e. a spreadsheet with a fault that actually has a quantitative impact. In the above example, if summing up B4 twice was not intended and B4 contained a non-zero value, the resulting value in C1 would very likely be wrong (if not hidden by another fault, that is where several faults balance each other and the result is still correct).
\item [Problem] refers to a negative impact in reality due to a failure in a spreadsheet. In the example above, if the wrong business decision was taken due to B4 being twice in the sum, this would be a problem.
\end{description}

The arrows in Figure \ref{fig:taxonomy} mean that there is a \enquote{can cause} relation between the anomalies, i.e. an error can cause a defect, a fault can cause a failure and a failure can cause a problem.

\subsection{Spreadsheet Inspections}

The presence of anomalies is an indicator of inadequate quality. However, software can only have reasonable quality because achieving \enquote{perfect} quality is not possible \cite{liggesmeyer2009software}. According to Fr\"uhauf et al., the usual way to achieve reasonable quality is by taking organizational, constructive and analytical steps. For traditional software, this means that a \enquote{systematic approach using proven software construction principles has to be followed}. The purpose of inspections is to \enquote{detect deviances from these principles} while \enquote{organizational steps provide the basic environment to do that} \cite[p. 20]{fruhauf2007software}.

Unlike professionally crafted traditional software, spreadsheets are usually developed unsystematically, not according to proven principles and by laymen. Therefore, it is not surprising that a meta study by Panko found that on average 94\% of the spreadsheets inspected in several studies contained anomalies \cite{panko2006spreadsheets}. Hence, the goal of diligent spreadsheet inspections is not to \enquote{inspect-in} adequate quality but to detect dangerous anomalies and to take such spreadsheets offline until they are sanitized.

In order to detect anomalies in spreadsheets systematically, the spreadsheet has to be inspected. To issue an inspection, test rules are needed. Test rules describe how to test a spreadsheet. There is a huge spectrum of possibilities to specify and execute test rules. In previous work \cite{kulesz2014integrating}, we reviewed both spreadsheet inspection approaches typically encountered in practice as well as inspection approaches proposed in literature for their underlying test rules. We identified three non-disjunct classes:

\begin{description}
\item [Fully automated approaches] already contain test rules or derive them from the spreadsheets to be inspected. The whole inspection is done by a tool.
\item [Partly automated approaches] rely on the participation of users for specifying test rules. The execution of the inspection is done by a tool.
\item [Manual approaches] only provide general test instructions such as a workflow or checklists. The execution of the inspection is done by human experts (studies indicate that teams detect more defects than individuals and that experienced practitioners perform only about 16 -- 23\% better than untrained students \cite{bishop2007empirical}). The use of tools is optional -- tools in this area support human experts, e.g. by providing interactive checklists or a visualization of the spreadsheet which is adjusted for inspection purposes such as \cite{hermans2013analyzing} or \cite{kankuzi2014domain}.
\end{description}

Each of these classes has its pros and cons. In general, the higher the level of automation, the lower the effort for executing inspections --  while, on the other hand, the more humans are involved, the more the effort increases.

However, apart from highly domain-specific ontology-based approaches such as \cite{kohlhase2009compensating}, fully automated approaches have limited detection capacities \cite{zhang2016effective} and have not been very successful in detecting faults because they are mostly blind to semantic incorrectness \cite{panko1999applying}. Furthermore, the effects of design and coding rules for spreadsheets are mostly unknown \cite{kulesz2012investigating} and anomaly patterns used in automated detection techniques can lead to the reporting of around 90\% false positive findings \cite{cunha2012towards}. Thus, we are convinced that relying on fully automated inspection approaches is not sufficient.

\subsection{Compliance Systems}

As organizations are forced by law to \enquote{prove} that they took actions to verify the correctness of their spreadsheets, they often deploy so-called compliance systems. Compliance systems work as follows: They inventory and monitor all spreadsheets in an organization and execute fully automated inspections on them. The findings are then presented to the management or the affected users for sanitization.

By using compliance systems, organizations are able to comply with regulatory requirements. We doubt, however, whether the relevant quality attributes of the inspected spreadsheets receive the appropriate care. While some imperfections can be addressed in this way, fully automated inspection approaches have severe limitations when it comes to detecting faults and failures (as discussed previously). Furthermore, studies in end-user environments show that enforcing unconvinced end-users to change their behavior only makes them find ways to circumvent regulations \cite{nardi1993small, panko2012end}.

\section{Spreadsheet Guardian}
\label{sec:spreadsheet_guardian}

If one demanded collaborating spreadsheet users to adapt to the working habits of disciplined software engineers, the spreadsheet users would have to inspect their spreadsheets after each development activity before starting any filling activities. This would be inconvenient for them as the spreadsheet paradigm does not distinguish between these activities. Also, it is unlikely that typical spreadsheet users who are not aware of the necessity of taking quality assurance steps (due to the  overconfidence problem) would accept such a model.

Viewing the whole situation from the perspective of a spreadsheet user reveals the following picture: The spreadsheet user gets an undocumented and potentially faulty spreadsheet and is asked to view it, fill in data or even develop it further. Even if the spreadsheet user was convinced of the necessity to inspect (and eventually fix) the spreadsheet first, the user would have a hard time deriving the test rules for the inspection. Most annoyingly, all the effort for deriving the test rules would be lost in the long run if just one of the subsequent users  in the chain of collaborators did not care about inspections. After a short time, the spreadsheet would have degenerated back into its original (undocumented and potentially faulty) state.

This is the starting point of our idea for a different approach: Instead of trying to persuade the bulk of spreadsheet users to do inspections or even enforcing them, we separate the activities of providing a test specification, executing inspections and analyzing findings from inspections. The idea behind this is to use only the \enquote{inspection-convinced} spreadsheet users for providing test rules while making the harvest of this effort (findings about violations of the test rules) available to all spreadsheet users -- including the ones not persuaded of benefits from doing inspections. Therefore, we ask the following research question:

\begin{leftbar}
\noindent \textit{How effective can automated spreadsheet inspections be if they are specified only by few spreadsheet users but executed by many?}
\end{leftbar}

To provide a theoretical framework for our approach, we developed an approach named \enquote{Spreadsheet Guardian}. Spreadsheet Guardian models inspections as a relation between producers, processors and consumers:

\begin{description}
\item [Test rule producers] are spreadsheet users who specify test rules.
\item [Test rule processors] are machines that execute spreadsheet inspections according to given test rules and compute findings (typically, there is just one such machine).
\item [Findings consumers] are spreadsheet users who receive findings computed by test rule processors.
\end{description}

The interaction between producers, processors and consumers works as follows:

\begin{itemize}
\item The test rule producers specify test rules using a tool that supports several specification techniques (for more details see section \ref{sec:implementation}).
\item The specified test rules remain in the spreadsheet, even if it is shared with other users who do development activities without having the tool installed.
\item The execution of spreadsheet inspections is done continuously and in the background by the test rule processor, while spreadsheet users just work with the spreadsheet as usual, i.e. carrying out filling or development activities.
\item After the test rule processor finishes the execution of the test rules, the spreadsheet user is presented with the findings directly in the spreadsheet environment (e.g. in Microsoft Excel). This way, spreadsheet users become findings consumers without any effort on their part.
\end{itemize}

We claim that Spreadsheet Guardian provides a unique combination of advantages over existing approaches. To make them clear, we describe our theory in terms of constructs, propositions, explanations and scope in Table \ref{tab:propositionstable} as following the theory schema by Sj{\o}berg et al. \cite{sjoberg2008building}.

\input{table1.inc}

\section{Implementation}
\label{sec:implementation}

Since 2011, the first author has supervised ten student theses and other projects in which a tool prototype named \enquote{Spreadsheet Inspection Framework} (SIF) was developed to implement Spreadsheet Guardian. SIF and its capabilities have already been described in previous work \cite{kulesz2014integrating}. Therefore, this section covers SIF only briefly. 

SIF consists of two components -- an analytical core and a front-end. The front-end has been implemented as an add-in for Microsoft Excel, thus it provides a tight integration with the environment typical spreadsheet users work in. Both components are free software and available on the web\footnote{http://www.spreadsheet-inspection.org}.

SIF is a framework which provides a common umbrella for experimenting with various spreadsheet inspection techniques. It also supports a total of eight fully automated inspection techniques which allow users to scan spreadsheets both for stylistic issues (e.g., the reading direction of formulas, the presence of constants in formulas) as well as for fault patterns (e.g., references to empty cells, repeated references to the same cell as discussed in the example of section \ref{sec:background}). Apart from technical aspects (e.g., SIF's analytical core is written in Java while the front-end is written in C\#), SIF distinguishes itself from similar tools by three key features: (i) its capabilities for reporting and managing findings, (ii) its ability to run inspections continuously in the background and (iii) its support for two partly automated inspection techniques.

\begin{figure*}[h]
\centering
\includegraphics[width=1\linewidth]{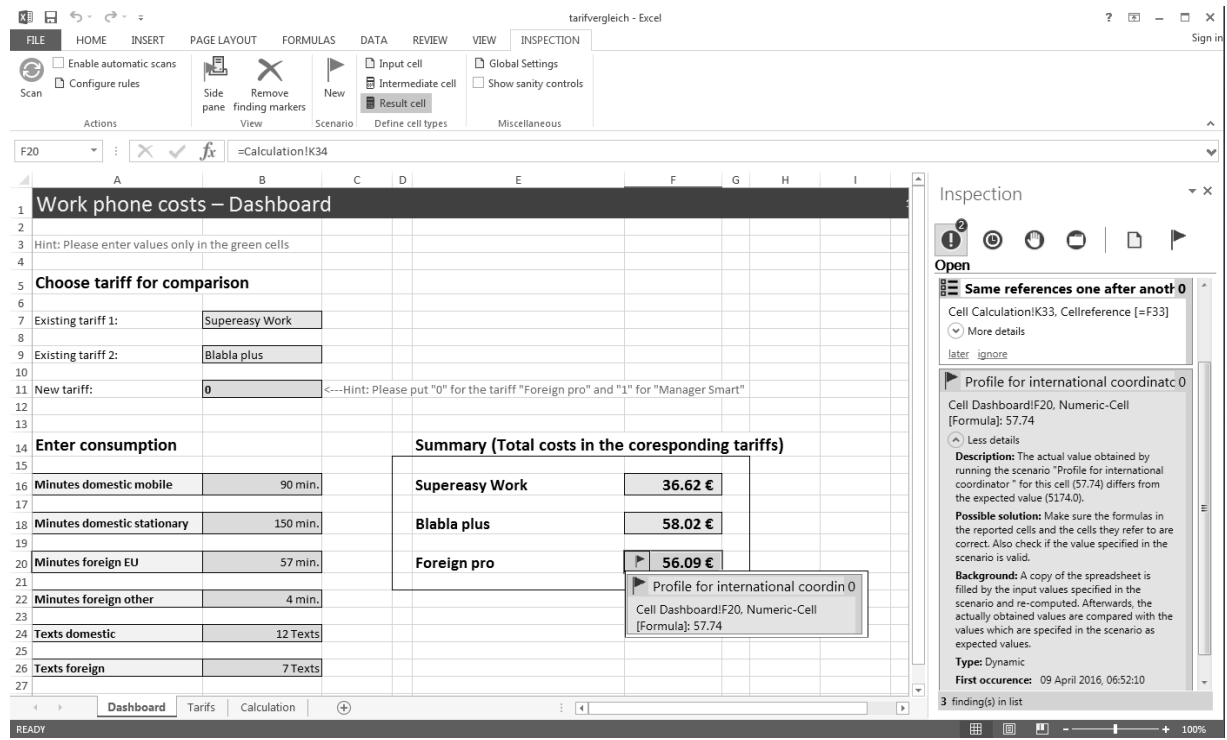}
\caption{Spreadsheet Inspection Framework's Microsoft Excel add-in reports findings in a spreadsheet}
\label{fig:screenshot}
\end{figure*}

As illustrated in Figure \ref{fig:screenshot}, SIF's front-end visualizes findings with marker icons directly in the spreadsheet as well as in a synchronized list in a side pane. This way, spreadsheet users are not distracted from their usual workflow but are still informed discretly about new findings (like in a modern IDE for traditional programming). Additionally, users can flag findings as false positives (so they do not reappear on subsequent inspections) or hold off on them.

SIF provides a so-called \enquote{live mode} which allows users to run inspections continuously in the background. A first user study from previous work indicates that this mechanism is generally acceptable and disruptive only on a low level for spreadsheet users' workflows \cite{kulesz2015live}.

Apart from providing the ability to use generic fully automated testing techniques (which are based on hard-coded test rules), SIF supports two unique inspection techniques which are partly automated and follow the approach proposed by Spreadsheet Guardian:

\begin{description}

\item [Test scenario technique] is a technique that allows spreadsheet users to specify their own test rules as follows: In the first step, the user marks corresponding cells in the spreadsheet as input cells, intermediate cells or output cells. Then the user specifies a set of values for the input cells and expected values or ranges of values for intermediate and output cells. SIF saves this set of values as a so-called \enquote{test scenario} as well as the marked cells in a separate region of the spreadsheet, so this data neither is lost nor gets in the way when other users change the spreadsheet (even if they do not use SIF at all). Basically, SIF's test scenario technique ports ideas from automated unit tests to spreadsheets. However, test scenarios for small spreadsheets are typically not split up into units but tested in an end-to-end fashion (input to output cells) even though the technique would be suitable for doing unit tests as well.

\item [Advanced data validation rules] allow users to specify test rules for the purpose of data validation. Unlike the restricted mechanisms that are provided in common spreadsheet execution environments, SIF's advanced data validation rules can be composed of multiple conditions to be checked. For instance, it can be specified that if an entry in column A starts with the letters \enquote{foo}, the corresponding cell in the same row but in a different column must contain a 10-digit number followed by the letters \enquote{bar}.

\end{description}

A typical workflow involving SIF might be as follows:

\begin{itemize}
	\item User 1 creates a test specification in SIF's front-end. In case of the test scenario technique, User 1 would do this by marking cells (as input/intermediate/output cells), providing actual values for input cells and expected values for intermediate/output cells.
	\item SIF's front-end saves the test specification together with the spreadsheet.
	\item (Optional) User 1 passes the spreadsheet to User 2. If this does not happen, User 2 and User 1 will be the same person in the following steps.
	\item User 2 changes any formula in the spreadsheet, triggering an inspection.
	\item SIF's front-end sends an inspection request to SIF's backend. The request includes the test specification and a pointer to the actual spreadsheet file.
	\item SIF's backend opens a copy of the spreadsheet file and executes the tests. Therefore, when using the test scenario technique, it populates all input values and triggers a recalculation of the spreadsheet.
	\item SIF's backend computes a list of findings containing any violations of the test specification. In case of the test scenario technique, it does so by comparing the actual values in the output cells with those stated in the test specification.
	\item SIF's backend sends the list of findings back to SIF's front-end.
	\item SIF's front-end visualizes the received findings.
	\item User 2 is now notified about the findings and can react to them (e.g. fix them, postpone them, ignore them, update the test specification in case it is outdated etc.).
\end{itemize}

As spreadsheets typically are directly connected to the reality they pre- or describe (Lehman's laws of software evolution \cite{lehman1980programs} would classify spreadsheets as P-programs or E-programs), it is very common that rows and columns are added or removed during maintenance. Also, worksheets are often added, removed, moved or renamed. Even in cases where these additions or removals are not done for cells where users specified test rules using one of SIF's techniques, these cells might be relocated: For instance, if a user specified that cell B12 is an output cell for a test scenario, inserting a new row between rows D and E would move the output cell to B13. To make SIF's techniques more resistant to such typical manipulations, SIF's implementation does not save absolute locations of the cells. Instead, it assigns hidden names to cells used in test scenarios and advanced data validation rules and refers to the cells only by these names. This way, test rules specified by SIF often can be left 
untouched even if structural changes in the spreadsheets are made.

\section{Evaluation Planning}

Spreadsheet Guardian is a conceptual approach and, thus, cannot be evaluated in practice without an implementation. SIF provides an implementation that fits this concept. Hence, we used SIF to evaluate Spreadsheet Guardian. Since the implementation of the test scenario technique in SIF was more mature than the implementation of the advanced data validation rules, we decided to do the evaluation based on this technique.

In previous work, we had evaluated both SIF's test scenario technique and advanced data validation rules technique only in experiments on a tiny scale. This was not sufficient to draw conclusions about the overall concept of Spreadsheet Guardian. Therefore, we conducted a comprehensive evaluation based on three experiments with different populations (a justification will be provided in section \ref{subsec:participants}).

\subsection{Goals}

Our main research objective was to evaluate Spreadsheet Guardian's suitability for the maintenance of spreadsheets. Therefore, we formulated the following goals for the evaluation:

\begin{itemize}
	\item Goal 1: \emph{Analyze} spreadsheet users learning Spreadsheet Guardian and creating test scenarios using SIF\\
	\emph{For the purpose of} evaluating the effectiveness of spreadsheet users specifying test scenarios ($\rightarrow$ P1)\\
	\emph{With respect to} the amount of correctly created test scenarios, the time required by users to specify the test scenarios and the perceived complexity of the maintenance tasks.
	
	\item Goal 2: \emph{Analyze} spreadsheet users maintaining pristine spreadsheets versus spreadsheets protected by Spreadsheet Guardian using SIF\\
	\emph{For the purpose of} comparing the effectiveness of the created test scenarios ($\rightarrow$ P2)\\
	\emph{With respect to} the actual correctness of the spreadsheets.
\end{itemize}

\subsection{Hypotheses}

We formulated three hypotheses (the first two target proposition P1/Goal1, the last targets P2/Goal2 (Table \ref{tab:propositionstable})) for evaluating the aspects of learnability, added complexity and correctness.

\subsubsection{$H_1$ - Feasibility}

\begin{itemize}
	
	\item[$H_1$] Spreadsheet users specify at least one effective test scenario correctly using SIF.
	\item[$H_{1_0}$] Spreadsheet users cannot specify any test scenario correctly using SIF.
	
\end{itemize}

(A test scenario is specified correctly if (i) it refers to all cells that provide data needed for the tested calculation as input cells, and (ii) it refers to all cells that show results depending on the tested formula as output cells, and (iii) it provides reasonable values for the input cells and the tested scenario and, (iv) the expected values for the output cells are correct regarding the used formulas and the given input values.)

\subsubsection{$H_2$ - Added complexity}

\begin{itemize}
	\item[$H_2$] Spreadsheet users perceive maintenance tasks to be more complex using SIF.
	\item[$H_{2_0}$] 	There is no difference in the perceived complexity by spreadsheet users using or not using SIF.
\end{itemize}	

(If using SIF adds too much burden on users maintaining spreadsheets, applying Spreadsheet Guardian could be infeasible for them.)

\subsubsection{$H_3$ - Correctness}

\begin{itemize}
	\item[$H_3$] After maintenance, spreadsheets not protected by Spreadsheet Guardian contain more faulty output cells than those protected by Spreadsheet Guardian.
	\item[$H_{3_0}$] After maintenance, there is no difference in the number of faulty output cells between spreadsheets not protected by Spreadsheet Guardian and those protected by Spreadsheet Guardian.
\end{itemize}

\subsection{Procedure}

\begin{figure}[h]
	\centering
	\subcaptionbox*{}%
	[.4\linewidth]{\includegraphics[scale=0.6]{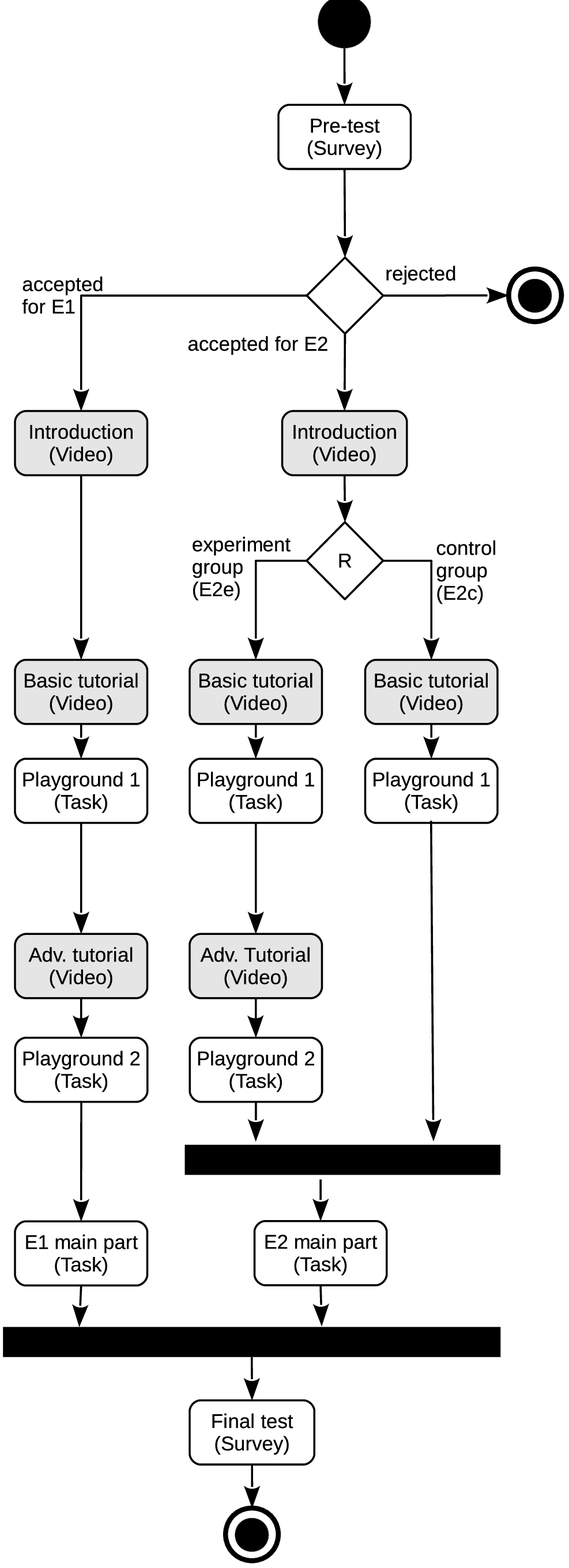}}
	\subcaptionbox*{}
	[.4\linewidth]{\includegraphics[scale=0.6]{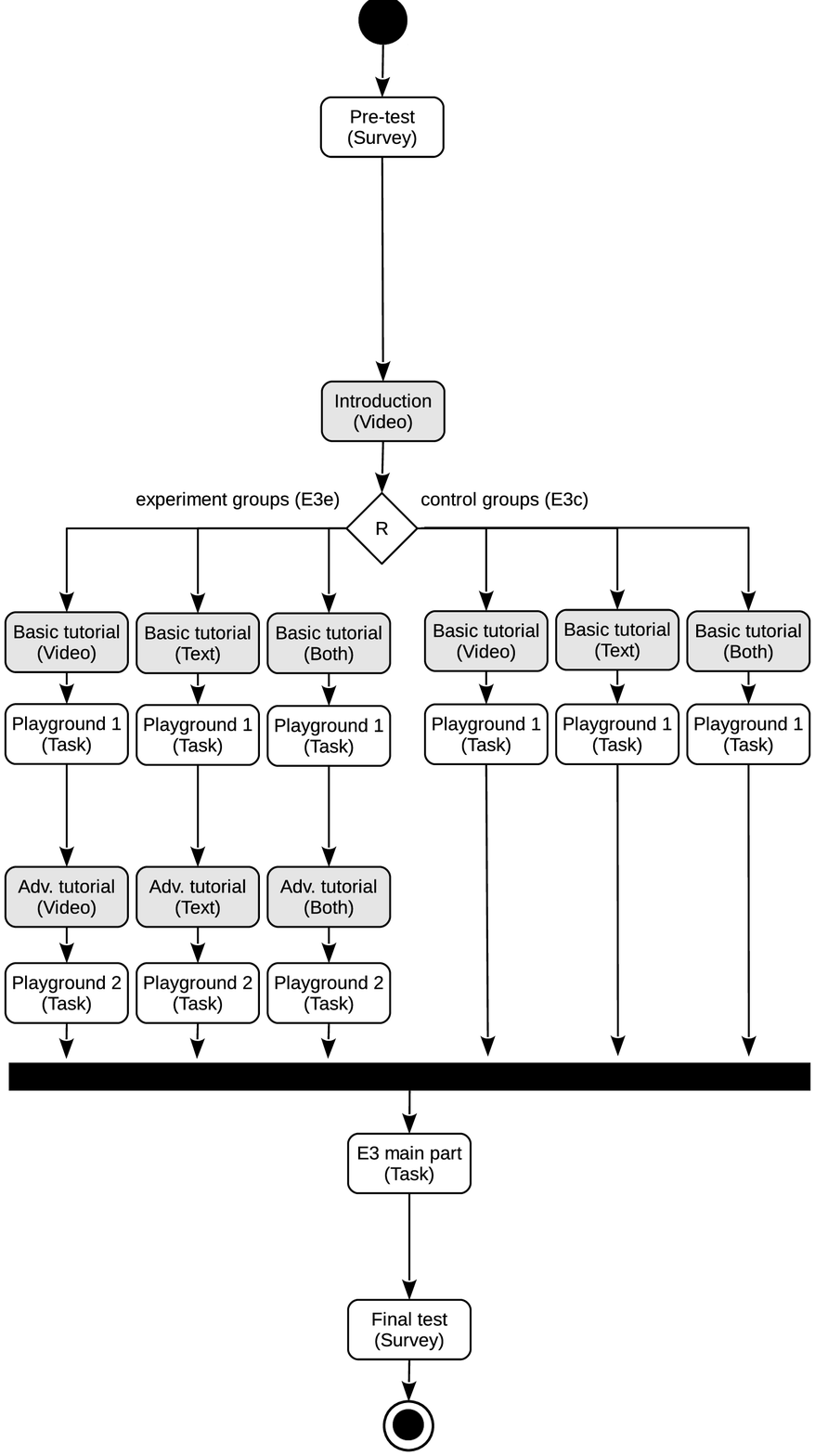}}
	\caption{Procedure of the experiments}
	\label{fig:procedure}
\end{figure}

We designed three experiments: E1, E2 and E3. In E1, we wanted to simulate the \enquote{sender side} where spreadsheet users protect their spreadsheets using SIF before sharing them. In E2, we wanted to simulate the \enquote{receiver side} where spreadsheet users carry out maintenance activities on spreadsheets \enquote{unfamiliar} to them. The experiment group was was given spreadsheets protected by Spreadsheet Guardian while the control group had to maintain \enquote{unprotected} spreadsheets. E3 was a moderately modified replication of E2 using a different group of participants. The planned procedure is illustrated in Figure \ref{fig:procedure}.

The procedure for E1 and E2 was planned as follows: After a pre-test (which was the same for all candidates), we computed a \enquote{suitability score} for each candidate and decided whether the candidate should be used for E1 (score: more than 60 points), E2 (score: more than 80 points) or be rejected (score: less than 60 points). The suitability score was calculated as a weighted sum over some answers from the questionnaire (the relevant questions are printed in Table \ref{tab:suitability_score}), however, the questionnaire contained more questions but not all of them were relevant for computing the score). While candidates for E1 were planned to be assigned directly,  candidates for E2 were planned to be randomly split into an experiment group (E2e) and a control group (E2c). Because our experiment was about the maintenance of spreadsheets, we planned to reject novices who typically only carry out filling activities but never develop or change formulas.

\input{table2.inc}

Experiment E3 was planned as an enriched variant of E2, differing only in two aspects:

\begin{itemize}
	\item The pre-test was planned to be issued only to gather data about the participants but not for rejecting \enquote{unsuitable} candidates.
	\item The participants were supposed to learn SIF with different (but content-equivalent) tutorial types instead of only video tutorials. We implemented this variation to investigate the effectiveness and efficiency of different tutorial types. Explaining the details behind this is beyond the scope of this paper, but we reported the results already in \cite{kafer2016best}.
\end{itemize}

During the experiments, all participants received the same introduction and the same basic training. However, participants of E1, E2e and E3e received additional training which participants of E2c and E3c did not. After the training, the participants were asked to solve the main part of their experiment. The main part of E2 and E3 was the same while E1 had a completely different main part. At the end, all participants were asked to complete a small survey that differed between the experiments only in small nuances (to keep it simple, this difference is not reflected in the aforementioned figures of the procedure).

\subsection{Participants}
\label{subsec:participants}

To acquire participants for E1 and E2, the first author posted messages in social media channels, contacted friends directly and distributed self-designed paper brochures on various occasions (e.g. sport courses, birthdays and house-warming parties). We succeeded to attract a total of 48 candidates for E1 and E2 but had to reject 7 unsuitable candidates and 4 suitable candidates broke up contact. Not counting our 8 pilots, we had a total of 29 participants who finished the experiments for E1 and E2. All participants of E1 and E2 were rewarded with 10 Euro and a chocolate bar for their participation.

The goal of our pre-test in E1/E2 was to triage our population of candidates for E1/E2 into spreadsheet beginners, intermediates and advanced users. This was necessary to match the premises of our research questions and to make sure that the level of difficulty was appropriate for the participants so they would neither feel underchallenged nor overstrained. Consequently, another goal of the pre-test in E1/E2 was to identify and reject candidates with a clear background in computer science or software engineering.

In contrast, the participants for E3 were acquired from an undergraduate software engineering course where participation in a study was mandatory. Thus, participants of E3 did not receive any monetary compensation for their effort. Apart from our study, the students could also choose from two alternative studies. In our advertisement, we stated that taking our experiment required basic spreadsheet skills. We used two pilots for E3 and succeeded to fill all our available slots to a total of 42 candidates. All of them showed up and took part in the experiments. 

\begin{figure}[h]
	\centering
	\subcaptionbox{Q: Which age group do you belong in?\label{fig:age}}%
	[.4\linewidth]{\includegraphics[width=0.4\linewidth]{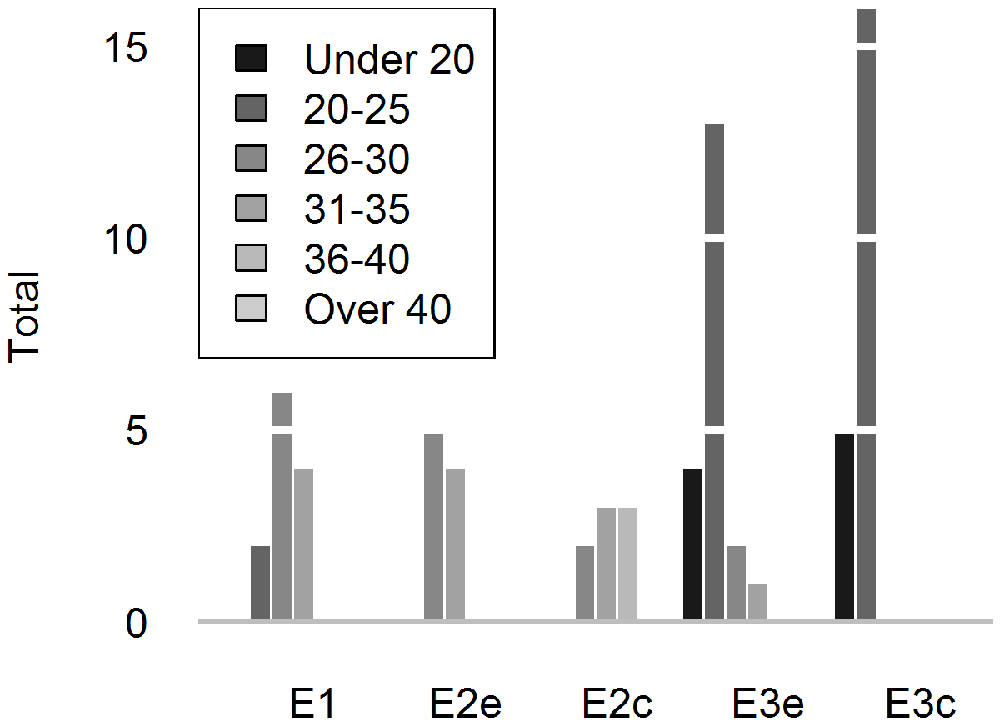}}
	\subcaptionbox{Q: What is your gender?\label{fig:males_females}}
	[.4\linewidth]{\includegraphics[width=0.4\linewidth]{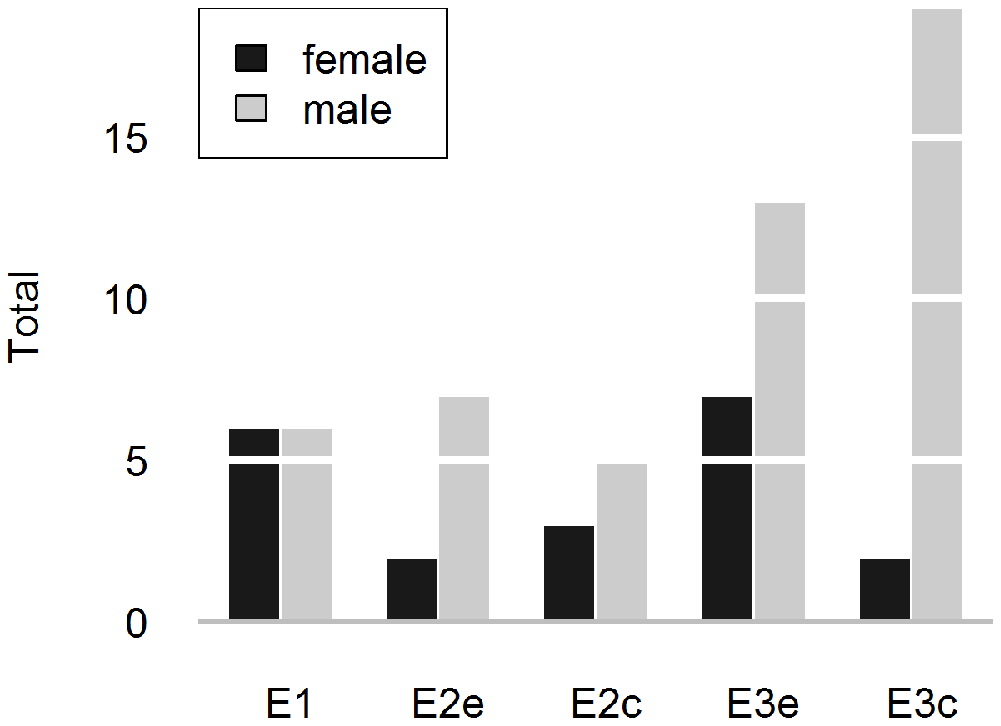}}
	\subcaptionbox{Q: How many years of professional work experience do you have?\label{fig:prof_exp}}
	[\linewidth]{\includegraphics[width=\linewidth]{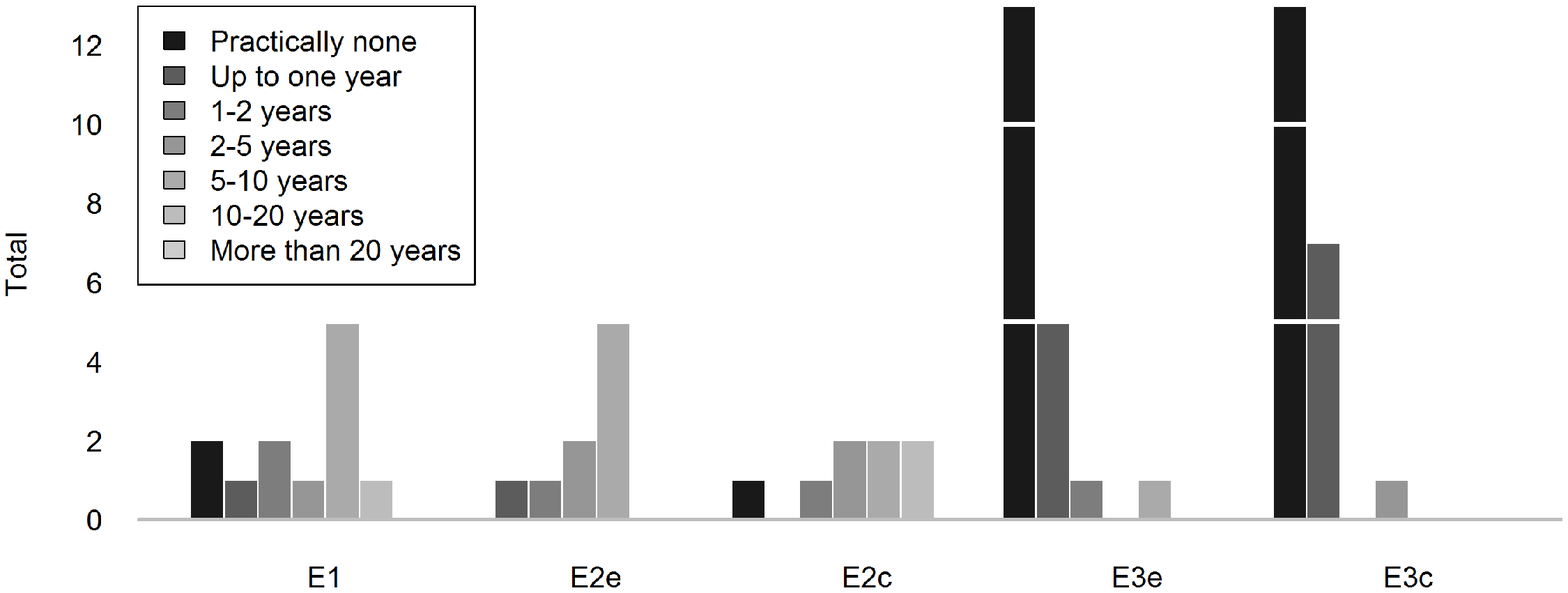}}
	\caption{Demographic data about the participants}
	\label{fig:demographic}
\end{figure}

We report demographic data of our samples in Figure \ref{fig:demographic}\footnote{The graphical representations in this work are based on the recommendations by Tufte~\cite{tufte}.}. When comparing participants from E2 and E3, it can be seen that the participants in E3 were younger and had less work experience. Since in E2 and E3 we randomly assigned participants to the control and experiment groups, the distribution of genders turned out to be rather unbalanced -- especially between E3e and E3c.

As can be seen, we used different populations in E1/E2 and E3. At first sight, one might argue that using \emph{solely} typical spreadsheet end-users would have been a better choice for targeting our goals -- and this was also our initial plan. However, after conducting E1 and E2 we learned that attracting a sufficient number of participants -- especially advanced spreadsheet users with some years of professional experience -- was very hard. Furthermore, finding suitable time slots for the participants who have a full-time job and/or who have family turned out to be very challenging. This forced us to find a balance between tightly controlling environment variables (like time, lighting conditions or external distraction factors) and offering flexibility to reach more participants (or to find participants at all).

We addressed this issue by allowing participants of E1/E2 to take the experiments off-site (in their homes) and in the evening or during the weekends (when they were tired after work or not in the mood for concentrating on hard work). However, to address the shortcomings of this compromise, we decided to slightly modify E2 and replicate it in a more controlled environment -- at the cost of running it with potentially less representative computer science students.

\subsection{Experimental Material}

Except for videos (which are available only upon request\confnote), all experimental material is available from our open data repository \cite{kulesz_daniel_2016_188896}. In the following, we describe the material briefly.

\subsubsection{Instructions}

All instructions during the experiments were given on paper sheets (11 pages for E1, 10 pages for E2e and E3e, 7 pages for E2c and E3c) in German\footnote{We translated the spreadsheets in the figures to ease understanding.}. During the experiments, the participants were asked to read the instructions and follow them. This involved watching video tutorials or reading text tutorials, applying the steps learned by doing tasks in spreadsheets and answering questions on paper. We reported more details about the experimental material in \cite{kafer2016best}.

\subsubsection{Pre-test}

The pre-test was implemented as an online survey. The survey had 21 questions and was loosely based on the survey used in the study of Baker et al. \cite{baker2006survey}. However, our variant was much shorter to make it completable in 5 to 10 minutes so we would not risk losing candidates for E1/E2. The questions were targeted at rating the experience of the participants and checking which spreadsheet activities they typically perform (e.g. just filling activities or also development).

For the actual triage after the pre-test, we developed a simple scoring sheet which computed a weighted sum based on the answers. 

\subsubsection{Training}

Understanding Spreadsheet Guardian and learning SIF requires some basic training. To provide adequate training to all participants and keep them motivated during the training phase, we employed a training concept with short \enquote{learn-and-apply} cycles: after watching recorded videos or reading text tutorials, the participants had to do small practical tasks in spreadsheets.

The basic training started with an introduction video (duration: 9m 3s) to spreadsheets, anomalies in spreadsheets and design rules for spreadsheets. To avoid unnatural behavior of the participants regarding our experiment questions, the video tried to give the participants the false impression that the experiment's goal was to measure effects of different spreadsheet design rules.

In the second part of the basic training we explained SIF's user interface in a video tutorial (duration: 5m 42s) or an adequate text tutorial. The tutorials showed how rules can be configured, how inspections can be started, how findings are reported and how they can be managed. After the tutorials, participants were asked to replicate the steps on a small \enquote{playground spreadsheet} (spielwiese.xlsx) that contained seeded faults.

The advanced training started with a video tutorial (duration: 10m 1s) or text tutorial that explained our test scenario technique and how to apply it in SIF. Again, the participants were asked to apply the technique to the playground spreadsheet afterwards. This involved marking cells, creating a test scenario, interpreting its results and finding the (seeded) cause for a failure.

\subsection{Tasks}

For the main tasks after the training, we invented a fictive setting where an existing spreadsheet for comparing mobile phone tariffs named \enquote{tarifvergleich.xlsx} had to be studied and maintained. To make this spreadsheet as realistic as possible, we asked an experienced spreadsheet user to produce the spreadsheet according to a given set of requirements. We then modified the spreadsheet's calculation worksheet to make it fit our experimental design and seeded two faults and one \enquote{evil} imperfection as shown in Figure \ref{fig:pristine}:

\begin{itemize}
\item The sum formulas in cells K33 and K34 added a value twice (like in the example in section \ref{sec:background}).
\item The reference in cell J34 was pointing to the wrong cell, i.e. to J29 instead of J30 (Excel's built-in mechanism fails to detect this).
\item The indexes for all VLOOKUP functions were stored in separate cells but with white color on white background, e.g.\ cell H28 contains a VLOOKUP function which reads the index from cell H25 that looks empty at first sight (we observed such practices in private inspections of spreadsheets; they are so common that they are explicitly mentioned in practitioner's literature on good and bad practices such as \cite{raffensperger2002art}).
\end{itemize}

Seeding defects is not as representative as using real-life defects. However, we did not find any real spreadsheets that would fit our experiment design. The main criteria here were that (i) the spreadsheets were domain-independent (i.e. previous knowledge in a specific domain like finance was not necessary), and (ii) the testing techniques in SIF were able to detect anomalies that could be seen as indicators for such defects. Therefore, we seeded defects that fulfilled these criteria. In addition, we selected only defects that were presented as common in literature about Best Practices for spreadsheets \cite{raffensperger2002art}\cite{o2005spreadsheet} and that we observed during earlier spreadsheet creation experiments with students \cite{kulesz2012investigating}.

Recently, Schmitz and Jannach published a corpus of real-world spreadsheet defects \cite{schmitz2016finding} from the Enron corpus \cite{hermans2015enron} that is based on the analysis of e-mail communication between collaborating employees at Enron. However, since the underlying spreadsheets are not domain-independent, they would have not been suitable for our experiment. After analyzing the particular defects from this corpus in detail, we conclude that many of them (such as \#1, \#2, \#3, \#4 or \#6) would have met the criteria for our experiment as well. Yet, some of the defects seeded by us even were present in the Enron error corpus \enquote{unintentionally}: Our playground spreadsheet had a logical fault (comparable to defects \#12, \#22, \#25 or \#29 from the Enron error corpus), while \enquote{tarifvergleich.xlsx} referred to the same cell twice (almost the same as defect \#23 from the Enron error corpus) and had a faulty cell reference (practically the same as \#30 from the Enron error corpus).

\subsubsection{Main Part of E1}

The main part of E1 consisted of four tasks. In the first task, the participants had to recompute the result for a given set of input values with the calculator of the operating system (not the spreadsheet). The second task was to create a new test scenario using SIF with the same values to show them that it would report the finding. In the third task, the participants were asked to look for the seeded defects in the spreadsheet and to try removing them so that the finding reported by the test scenario technique would be eliminated. They were given the hint to activate the fully automated inspection techniques (which gave hints for the first fault) but had to find the second fault manually. In the last task, they were asked to \enquote{invent} consumption values for a described cell phone user and to compute expected values using a calculator before using SIF to create and run a test scenario for it.

\subsubsection{Main Part of E2/E3}

Unlike in E1, in E2/E3 we wanted to simulate a situation on the \enquote{receiver side} of a user maintaining a spreadsheet. Thus, the focus here was less on corrective maintenance (tracing failures and finding faults) but on adaptive maintenance, i.e. developing the spreadsheet further by adding more features. 

To tackle the experiment question, we divided the participants into two groups -- an experiment group (E2e/E3e) and a control group (E2c/E3c). However, we think that it would have been too hard to isolate involved effects if we had one group with all the training and SIF's complete stack and one group with more or less nothing. Therefore, we decided to give both groups the same basic training and SIF's fully automated inspection techniques but to provide only the experiment group with SIF's test scenario technique. This way, both groups were using SIF but only the experiment group was using the approach proposed by Spreadsheet Guardian.

Both the experiment groups (E2e/E3e) and the control groups (E2c/E3c) received the same task and almost the same spreadsheet (\enquote{tarifvergleich.xlsx}). However, only the experiment groups' spreadsheet already had four test scenarios planted in. These test scenarios were the first four correct test scenarios that the participants of E1 had produced. We chose this approach to avoid artificial scenarios.

\begin{figure}[h]
	\centering
	\subcaptionbox{Pristine\label{fig:pristine}}%
	{\includegraphics[width=0.49\linewidth]{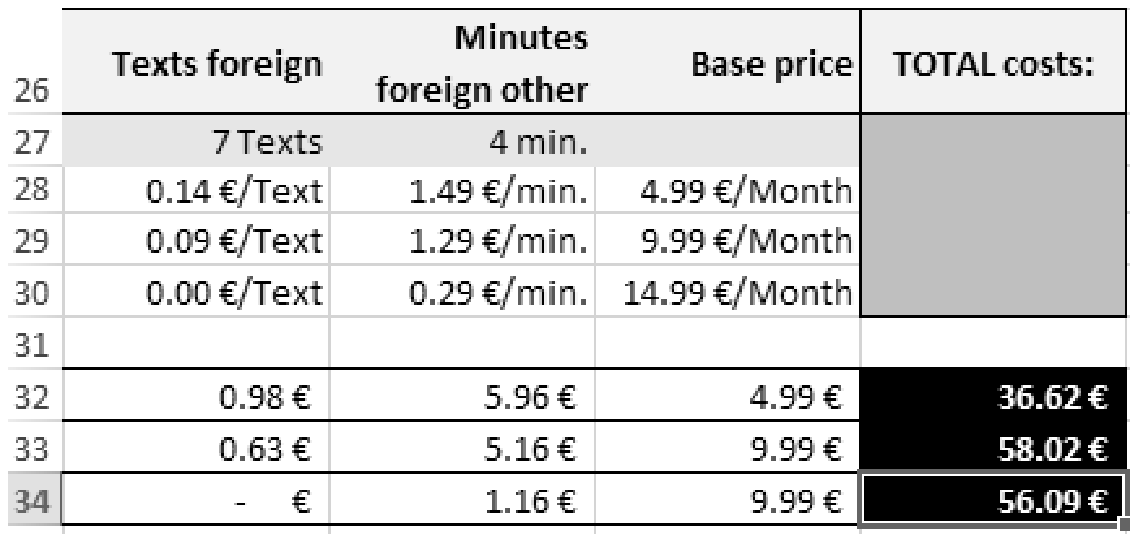}}
	\subcaptionbox{After erroneous maintenance\label{fig:maintained}}
	[.49\linewidth]{\includegraphics[width=0.49\linewidth]{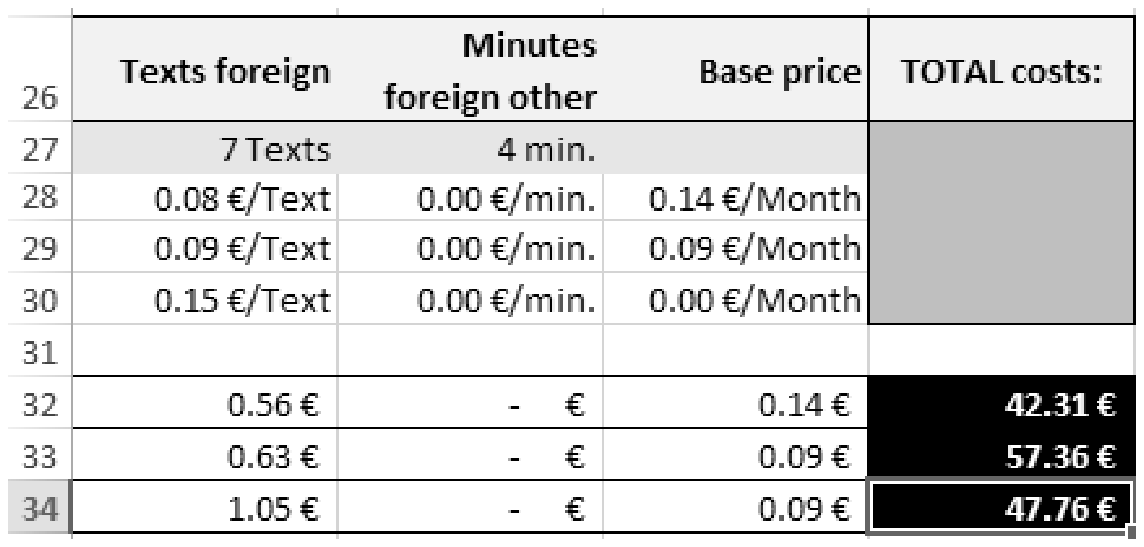}}
	\caption{ Excerpt of \enquote{tarifvergleich.xlsx}}
	\label{fig:experiment_sheets}
\end{figure}

The main part of E2/E3 had just three tasks. In the first task, the participants were asked to activate tests (including test scenarios and tests based on other test rules), run them and fix the reported findings. While participants of E2e/E3e were given concrete findings pointing to the second seeded fault, the participants of E2c/E3c were not given any hints  -- but we also did not expect them to detect it. In the second task, the participants were asked to add a new tariff and to include it in the comparison of the \enquote{Dashboard} worksheet.

The crux was the third task: here, we asked the participants to extend the spreadsheet by a new consumption category (\enquote{Texts network-internal}). The deliberately nasty trap here was that by inserting a new column in the tariffs worksheet, the hidden indexes used for the VLOOKUP functions in the Calculation worksheet would not be automatically updated -- resulting in obvious failures as shown in Figure \ref{fig:maintained}.

\subsubsection{Final Test (FT)}

The purpose of the final test was to investigate how our participants liked the experiment and our SIF tool, how confident they felt about their modifications being correct (only E2/E3) and to gather more data about their habits and background. This was done by a paper survey which consisted of 33 questions. The questions were a mix of questions adapted from the survey by Baker et al. \cite{baker2006survey} and our own.

\subsection{Analysis Procedure}

To judge the correctness of the \enquote{tarifvergleich.xlsx} spreadsheets in E2/E3, we designed eight simple test scenarios -- four for regression testing and four for testing the functionality the participants were supposed to add. To make it fair for the control groups, we treated output values as correct even if they did not fix the second seeded fault (which was hard to find without the hint given by SIF's test scenario technique).

We executed the tests manually (without the help of SIF) twice, i.e. we opened each spreadsheet, filled in the test scenario's input values and captured the result values. Our results were not normally distributed, therefore we decided to use a Wilcoxon signed-rank test for statistical analyses. Additionally, we computed Cohen's d for the effect size. To measure normal distribution as a t-test precondition we used a Shapiro-Wilk test.

\section{Evaluation Execution and Analysis}

All experiments of E1/E2 were done in a time frame of six weeks. The execution of E3 was done in two weeks. However, designing, preparing and piloting the experiments as well as evaluating the results took us about seven months.

\subsection{Instrumentation}

We did not want to breathe down the participants' necks during the experiments. Therefore, we watched the experiments over VNC connections from the experimenters' machines. In general, we assume that nowadays it is normal for spreadsheet users to have internet access and, therefore, an experiment feels more natural if participants are allowed to access the internet during experiments. However, the off-site setting in E1/E2 made it impossible to guarantee (equal) internet access to all participants. Thus we decided to do E1/E2 without it. This was not an issue for E3, so we allowed internet access there.

For E1 and E2 we did not impose strict time limits. We only asked the participants to reserve about 90 minutes for the experiment. Due to university regulations, for E3 we had to limit the available time to 120 minutes. When time was up, we told the participants so but they were free to continue the experiments if they wanted.

In general, we refrained from giving the participants hints and interfering with the experiments except for the following reasons:

\begin{itemize}
	\item A participant did not understand the instructions.
	\item A participant encountered one of SIF's known bugs.
	\item A participant was not sure what the parameters of the VLOOKUP function stand for. In such cases, we explained the function to participants of E1 and E2. For E3, we motivated the participants to do research on the internet themselves.
	\item A participant of E1, E2e or E3e had no idea how to find the defects (we gave the advice to enter the input values of a test scenario manually and continue searching).
\end{itemize}

\subsection{Observations}

We analyzed the collected data and looked for interesting findings. In the following, we will present a few selected results that are relevant especially regarding our experiment questions and hypotheses.

\subsubsection{Pre-test}

\begin{figure}[!h]
	\centering
	\includegraphics[width=1\linewidth]{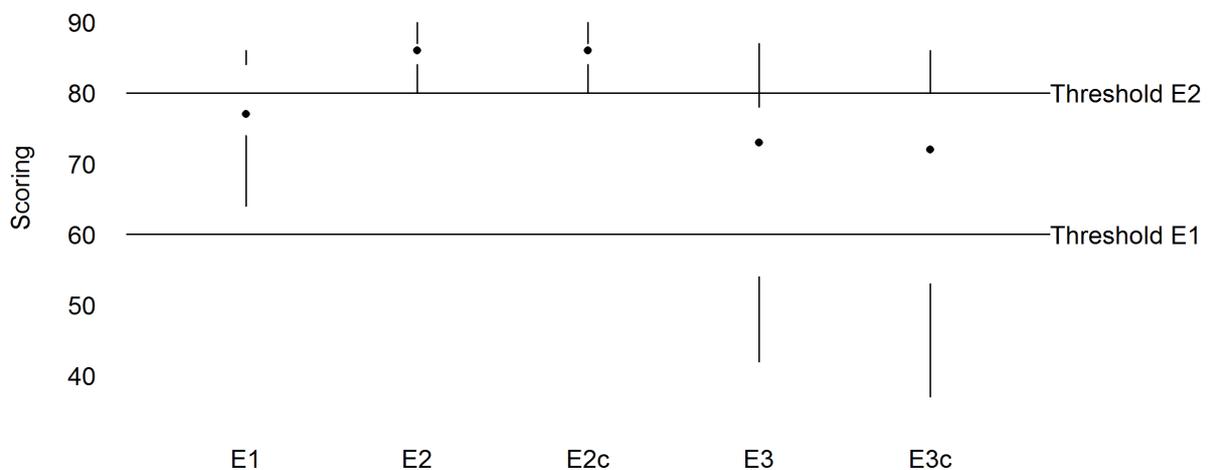}
	\caption{Participants' scores}
	\label{fig:scoringboxplot}
\end{figure}

Figure \ref{fig:scoringboxplot} shows the results of the triage. The horizontal lines mark the thresholds we used in E1/E2 to distinguish between suitable and unsuitable participants for E1 and E2. There is a significant difference between the scores of E2 and E3 (p=0.0005, d=0.6910). It can be seen that of the participants we accepted for E3, we would have rejected many for E2 (and some even for E1).

\subsubsection{Participants' Impressions}

\begin{figure}[h]
	\centering
	\subcaptionbox{Q: How did you like our study?\label{fig:like}}
	[.4\linewidth]{\includegraphics[width=0.4\linewidth]{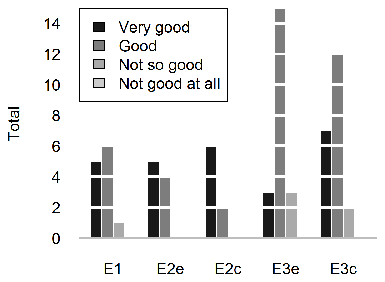}}
	\subcaptionbox{Q: How did you like our tool SIF?\label{fig:likeSIF}}
	[.4\linewidth]{\includegraphics[width=0.4\linewidth]{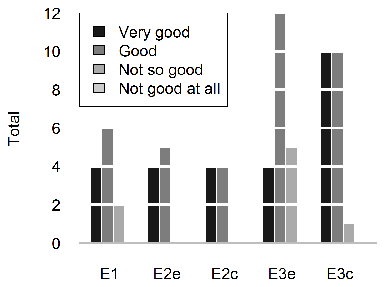}}
	\caption{Participants' impressions}
	\label{fig:wellness}
\end{figure}

In all experiments, only five participants stated that they did not like the study too much. The remaining participants stated that it was very good (26) or good (39) (Figure \ref{fig:like}). We think that participants who like a study are more likely to give their best during an experiment whereas participants who do not like a study tend to rush through it to get it over with. 
A similar picture can be drawn for SIF in general where eight participants stated not to like SIF too much, while the remaining participants stated that it was very good (26) or good (37) (Figure \ref{fig:likeSIF}).

\begin{figure}[h]
	\centering
	\subcaptionbox{Q: The visual appealing of the spreadsheet is well done.\label{fig:likeoptical}}
	[.49\linewidth]{\includegraphics[width=0.49\linewidth]{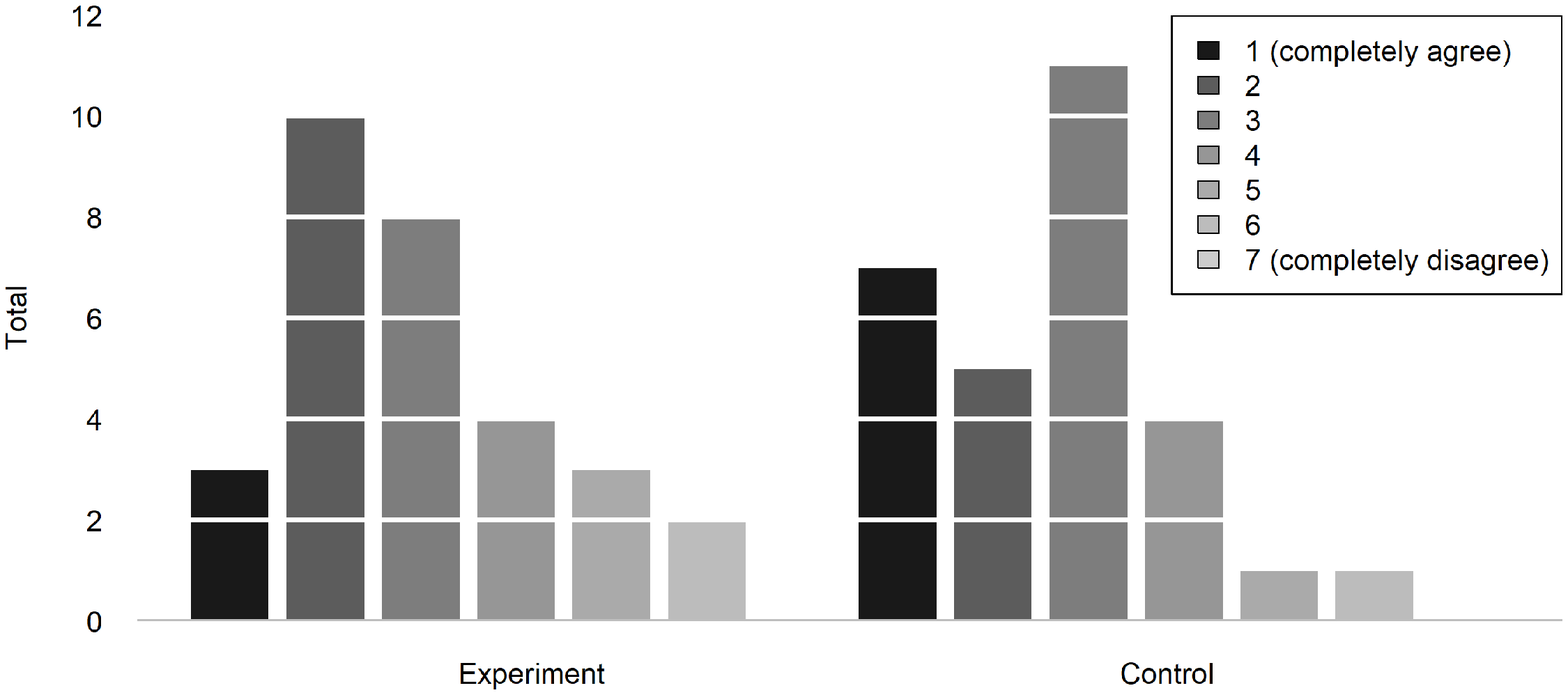}}
	\subcaptionbox{Q: The maintainability of the spreadsheet is good.\label{fig:likemaintainability}}
	[.49\linewidth]{\includegraphics[width=0.49\linewidth]{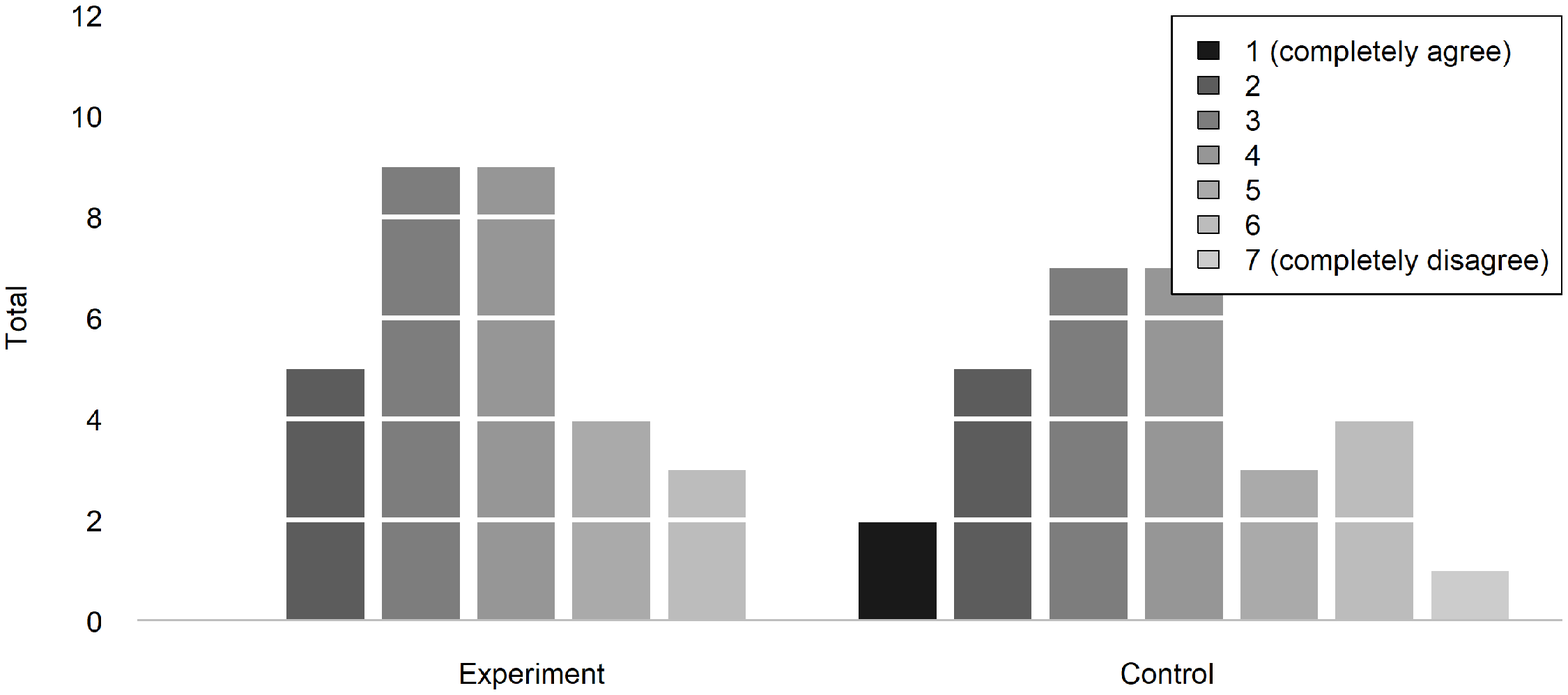}}
	\caption{Participants' ratings}
	\label{fig:spreadsheetratings}
\end{figure}

In addition, we asked the participants to rate the visual presentation and the maintainability of the spreadsheet they had to maintain in the main task. The results are shown in Figure \ref{fig:spreadsheetratings}. Regarding the visual presentation, both ratings are not distributed normally and show a slight bias to a good visual presentation. Totally, 44 participants agreed that the visual appealing was well done, 7 disagreed and 8 were not sure. 
Regarding the maintainability, the ratings of the control groups were normally distributed (14 liked the maintainability and 8 did not) whereas the ratings of the experiment groups were slightly in favor of a good maintenance (14 vs 7)

\subsubsection{Hypothesis $H_1$ -- Feasibility}

\begin{figure}[h]
	\centering
	\subcaptionbox{Completeness\label{fig:eingabe_alle}}%
	[.49\linewidth]{\includegraphics[width=0.49\linewidth]{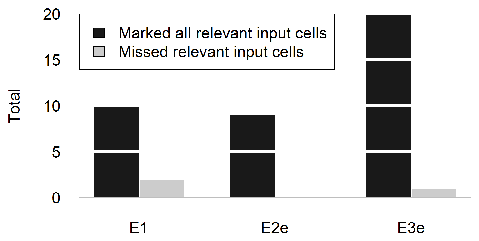}}
	\subcaptionbox{Correctness\label{fig:eingabe_unnoetige}}%
	[.49\linewidth]{\includegraphics[width=0.49\linewidth]{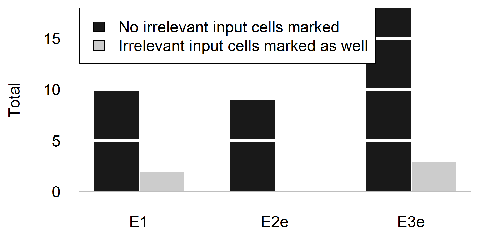}}
	
	\caption{Marking of the input cells}
	\label{fig:szen}
\end{figure}

For creating scenarios, it is necessary to first mark input, intermediate and output cells. We therefore analyzed whether the participants succeeded to do so. While \emph{all} participants successfully marked \emph{all} intermediate and output cells correctly, a small percentage of the participants missed some input cells or marked irrelevant cells as input cells (Figure \ref{fig:szen}). We did a t-test on E1 and E3e to see if a significant amount of participants selected all relevant input cells and it confirmed that they did ($p_{E1}$=0.001, $d_{E1}$=3.0277, $p_{E3e}$=4.277e-06, $d_{E3e}$=6.1721). The same is true for not marking irrelevant cells ($p_{E1}$= 0.001, $d_{E1}$=0.6055, $p_{E3e}$=1.233e-05, $d_{E3e}$=0.5634). We did not have to test E2e because all participants in this group marked all relevant input cells and no irrelevant ones.

Only creating test scenarios is not sufficient for identifying faults. A test scenario must also be effective to unveil a fault. This requires populating the marked cells with reasonable values when creating scenarios. The analysis of all created scenarios showed that \emph{all} participants provided reasonable values for all types of cells. In the cases where participants marked too many input cells, we observed that they simply filled irrelevant cells with zeros.
Therefore, we can reject $H_{1_0}$ and support $H_1$: spreadsheet users are able to specify at least one effective test scenario correctly using SIF.

\begin{figure}[h]
	\centering
	\subcaptionbox{Playground 1\label{fig:task1}}%
	[.49\linewidth]{\includegraphics[height=6cm]{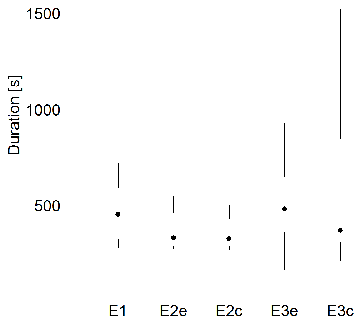}}
	\subcaptionbox{Playground 2\label{fig:ask1_sw}}
	[.49\linewidth]{\includegraphics[height=6cm]{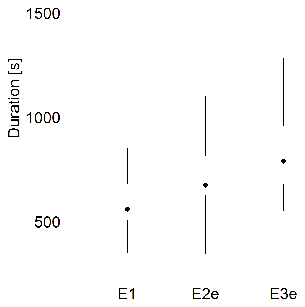}}
	\caption{Duration of the playground tasks in seconds}
	\label{fig:time}
\end{figure}

In addition to effectiveness, we also examined the efficiency of specifying test rules. To achieve this, we measured the duration of the playground tasks. We set 15 minutes as a reasonable time for playground 1 and 20 minutes for playground 2. Figure \ref{fig:time} shows the measured duration of the playground tasks.

As it can be seen, the participants solved the tasks quickly. A t-test confirmed that all groups on playground 1 were significantly faster than 15 minutes ((t) $p_{E1}$=0, $d_{E1}$=4.3115, (w) $p_{E2e}$=0.0089, $d_{E2e}$=4.8511, (t) $p_{E2c}$=0, $d_{E2c}$=5.6326, (w) $p_{E3e}$=0.0012, $d_{E3e}$=2.5828, (w) $p_{E3c}$=0.0089, $d_{E3c}$=1.8943). 

Also, for playground 2 a t-test confirmed that the participants were significantly faster than 20 minutes ((w) $p_{E1}$=0.001, $d_{E1}$=3.3167, (t) $p_{E2e}$=8.1019e-05, $d_{E2e}$= 4.5556, (w) $p_{E3e}$=0.0008, $d_{E3e}$=3.8861).

\subsubsection{Hypothesis $H_2$ -- Added complexity}

\begin{figure}[h]
	\centering
	\subcaptionbox{Playground 1\label{fig:suc_SW1}}%
	[.49\linewidth]{\includegraphics[width=0.49\linewidth]{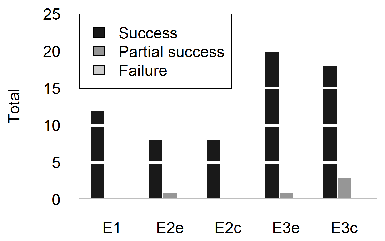}}
	\subcaptionbox{Playground 2\label{fig:suc_SW2}}%
	[.49\linewidth]{\includegraphics[width=0.49\linewidth]{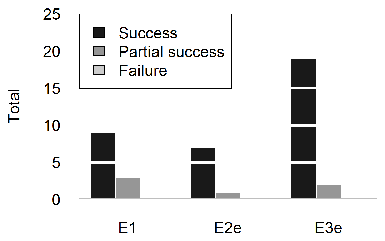}}
	
	\caption{Task-solving performance of the participants}
	\label{fig:suc}
\end{figure}

One could argue that if an activity is more complex, the success rate of applying it is lower. Therefore, we examined the success rate of both playgrounds. As Figure \ref{fig:suc} indicates, most participants were able to solve the tasks of both playgrounds.

\begin{figure}[h]
	\centering
	\includegraphics[width=\linewidth]{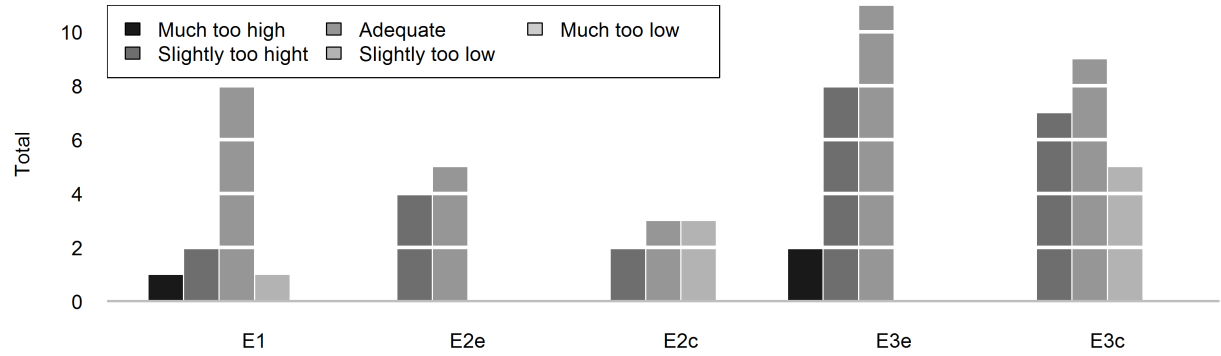}
	\caption{Q: For me, the difficulty level of the experiment was ...}
	\label{fig:difficulty}
\end{figure}

In addition to the tasks-solving success, we also analyzed the perceived difficulty to see if applying Spreadsheet Guardian using SIF added to this. As shown in Figure \ref{fig:difficulty}, the majority of the participants of E1 perceived the difficulty level as adequate. For both E2 and E3, some participants of the particular experiment group perceived the experiment to be slightly more difficult than in the particular control group. For E3, the experiment group perceived the experiment as significantly more difficult than the control group ((w) $p_{E3e-E3c}$=0.03654, d=0.658). For E2, there was no significant difference ((w) $p_{E2e-E2c}$=0.07189). 

After these analyses we cannot completely reject $H_{2_0}$: some of the participants perceived the application of Spreadsheet Guardian using SIF as significantly more complex while some did not.

\subsubsection{Hypothesis $H_3$ -- Correctness}

\begin{figure}[h]
	\centering
	\includegraphics[width=0.39\linewidth]{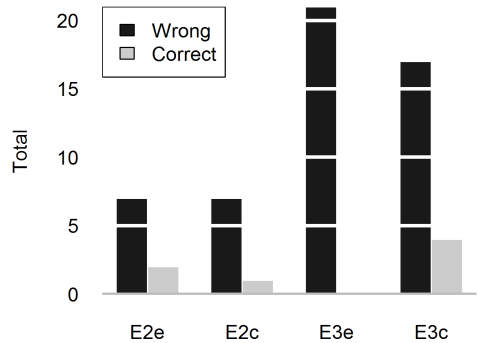}
	\caption{Correctness after maintenance activities in E2e/E3e and E2c/E3c}
	\label{fig:correctness}
\end{figure}

Figure \ref{fig:correctness} shows the results of the final task. As you can see, most participants could not solve the task in a way that produced correct results. In fact, in E3e not one single participant was able to do so. Nevertheless, in both E3 and E2 there was no significant difference of the results between the two groups ((w) $p_{E2}$=0.3313, (w) $p_{E3}$=0.9822). Therefore, we cannot reject $H_{3_0}$: at least in our sample, spreadsheets protected by Spreadsheet Guardian did not contain fewer faulty cells than \enquote{unprotected} spreadsheets. The fact that in one group nobody could solve the tasks is rather arbitrary in our view. As the number of participants who could solve the task is rather low, it is not surprising that in one group the number is zero. For what we know, there is no other explanation for this.

\subsubsection{Further Exploratory Analysis}

During our analysis, we noticed that some participants were more confident than others regarding the correctness of their spreadsheet after their modifications. Although this was covered neither in our initial propositions nor our hypotheses, we decided to explore this aspect further.

\begin{figure}[h]
	\centering
	\includegraphics[width=0.75\linewidth]{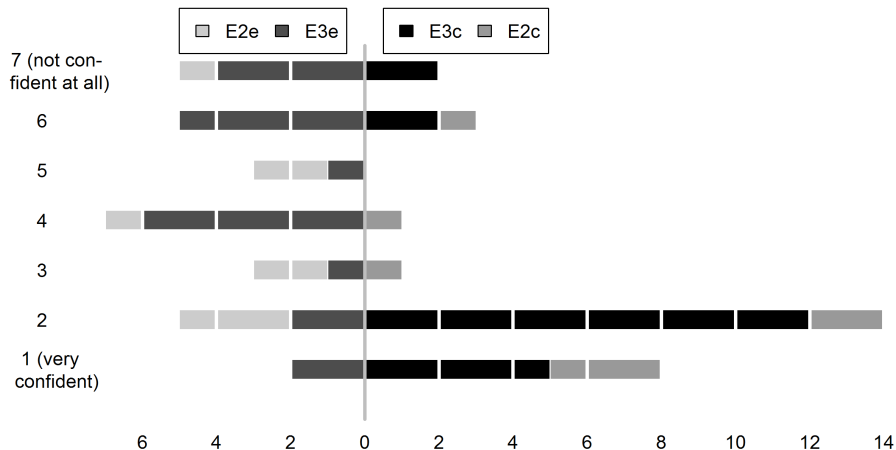}
	\caption{Q: How confident are you that your spreadsheet is correct?}
	\label{fig:confidence_correctness}
\end{figure}

In the final test, the participants had to state on a 7-point Likert scale how sure they were that the spreadsheet in the final task yielded correct results after their maintenance activities. Figure \ref{fig:confidence_correctness} shows their responses. As it can be seen, the participants of E2e and E2c were comparably confident whereas in E3 the participants of E3c were more confident than those of E3e.

\begin{figure}[h]
	\centering
	\subcaptionbox{E2e\label{fig:conf-Corr-E2}}
	[.49\linewidth]{\includegraphics[width=0.49\linewidth]{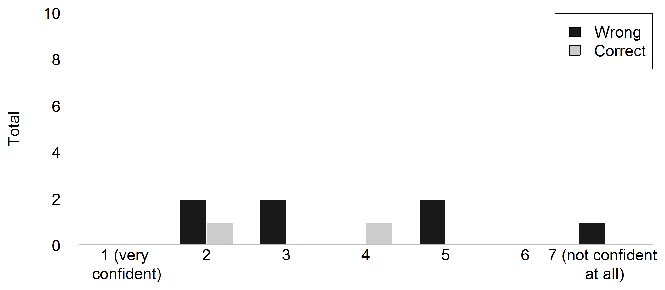}}
	\subcaptionbox{E2c\label{fig:conf-Corr-E2c}}
	[.49\linewidth]{\includegraphics[width=0.49\linewidth]{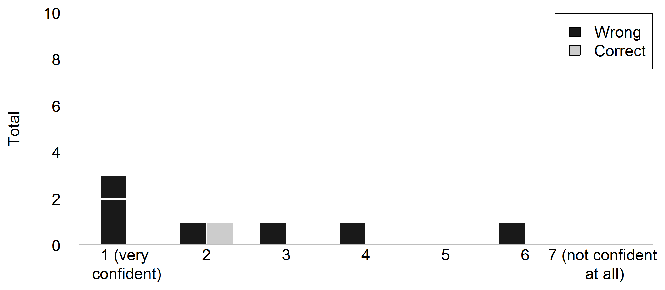}}
	\subcaptionbox{E3e\label{fig:conf-Corr-E3}}
	[.49\linewidth]{\includegraphics[width=0.49\linewidth]{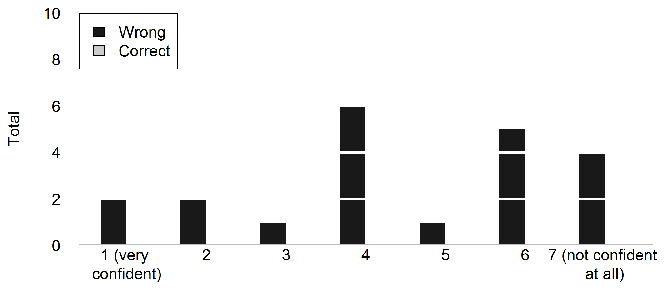}}
	\subcaptionbox{E3c\label{fig:conf-Corr-E3c}}
	[.49\linewidth]{\includegraphics[width=0.49\linewidth]{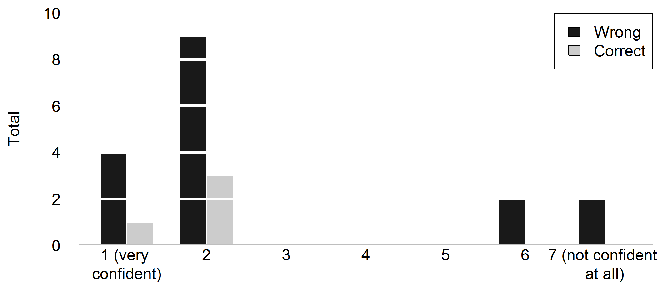}}
	\caption{Distribution of confidence across wrong and correct solutions}
	\label{fig:conf-corr}
\end{figure}

We also compared the actual correctness of the maintained spreadsheets with the participants' confidence about them being correct. As illustrated in Figure \ref{fig:conf-corr}, many participants of the control groups were very sure about the spreadsheets' correctness while in fact most were wrong. On the other hand, the self-confidence of the participants in the experiment groups was more balanced.

Further, we investigated the correctness of the participants' assessments. We counted an assessment as being correct if a participant's spreadsheet was correct and the participant was confident about it being correct. Also, when a participant stated a very low confidence and the spreadsheet was indeed incorrect, we evaluated this as a correct assignment. Likewise, wrong spreadsheets but participants being confident that it was correct as well as correct spreadsheets but participants being unconfident were counted as incorrect assessments. For this measure, we used the confident ratings (1, 2 and 3 on our Likert scale) and the unconfident ratings (5, 6 and 7 on our Likert scale) and disregarded the \enquote{unsure} rating (4 on our Likert scale).

Table \ref{tab:ratio} summarizes the results about the correctness of the participants' assessments. We also computed a ratio between correct and incorrect assessments for each group. The results show that participants of the experiment groups had much more realistic assessments about the correctness of their spreadsheets than participants of the control groups.

\input{table3.inc}

\subsection{Interpretation}

\subsubsection{Feasibility ($\rightarrow$ P1)}

As the results of our experiment E1 group show, applying Spreadsheet Guardian by specifying effective test rules using SIF's test scenario technique is easy and feasible even for regular spreadsheet users. The vast majority of our participants marked cells correctly and provided reasonable output values for test scenarios. Moreover, the time required to learn the test scenario technique is reasonably low -- two short tutorials were enough to teach all basics. Yet, one aspect that is partly surprising is that experienced Excel users seem to slightly outperform undergraduate computer science students here. This is interesting, as the computer science students are likely to be already familiar with the concept of unit testing (which is taught in the first semester of our program). This allows us to conclude that previous knowledge of the concept of unit testing is not required when it comes to learning and applying SIF's test scenario technique.

\subsubsection{Added Complexity ($\rightarrow$ P1)}

Typically, quality assurance is not for free. This is also true for applying Spreadsheet Guardian using SIF's test scenario technique, as it requires users to carry out additional activities which they would not typically do when maintaining spreadsheets. The key question we were interested in here was whether the additional effort stays within reasonable limits.

For spreadsheet users that do not work with spreadsheets on a daily basis and are already overstrained by using slightly advanced functions such as IF or VLOOKUP (like many in our E3 group), our maintenance tasks were definitely too difficult. Therefore it does not come as a big surprise that when asking them to do more (that is, additionally deal with Spreadsheet Guardian), they perceived the experiment as even more difficult. For those participants who were not able to solve the tasks, we analyzed the screencasts and came to the conclusion that it was not because they could not use SIF but because they lacked the knowledge how to solve the specific spreadsheet defect. Here, we assume that not knowing about the existence of a defect \enquote{feels easier} than being informed about a defect one is unable to solve.

Unlike these spreadsheet beginners, the participants of the E2 group did \emph{not} perceive using SIF's test scenario technique as significantly more complex than doing spreadsheet maintenance without it. This allows us to conclude that for this group the additional effort required for applying the technique is acceptable. However, the major challenge with this group is to convince them to actually apply the technique. Even if the effort for doing so is low for these users, they still need to see a benefit for spending this effort. But we are convinced that following a surprise-explain-reward strategy \cite{wilson2003harnessing} could help to bridge this gap.

\subsubsection{Effects on Correctness ($\rightarrow$ P2)}

When it comes to judging the usefulness of applying Spreadsheet Guardian by using SIF's testing technique for preserving correctness during maintenance, the results are not very clear. Be it with or without the test scenario technique -- the vast majority of our participants broke the spreadsheets' correctness during maintenance. While in E2 the users who employed the test scenario technique performed slightly better, in E3 it was the other way round. We think that it would require a sample with a higher maintenance success rate (and, thus, possibly easier tasks) to make profound judgments on this aspect. Further experimentation will be required to shed more light on this aspect.

\subsubsection{Effects on Overconfidence (not related to P1 or P2)}
While the correctness of spreadsheets is important, we consider having a realistic perception about the correctness to be even more important. When comparing the results from E2/E3 between the experiment and control groups, we were really stunned that the \enquote{invisible gorilla} effect \cite{chabris2011invisible} was much stronger than we had expected -- we previously thought that most participants would recognize at first sight that something must be terribly wrong with the numbers in the base price column (see Figure \ref{fig:experiment_sheets}).

From the results, we conclude that just deploying a tool with only fully automated inspection approaches might lead to a false perception of correctness (if no findings are reported by such a tool, users might be even more confident that the spreadsheet is correct). In contrast, applying Spreadsheet Guardian by using SIF's test scenario technique seems to successfully fight overconfidence. Actually seeing that there are still findings in a spreadsheet will, of course, leads to less confidence that the spreadsheet is correct, otherwise the findings would be useless. And maybe realizing (while working with SIF) that there actually were findings that were not obvious to the naked eye will lead to less overconfidence as well. This encourages us to extend our original theory (described in section \ref{sec:spreadsheet_guardian}) with a new Proposition P3 as outlined in Table \ref{tab:proposition_extend}.

\input{table4.inc}

Another plausible explanation for the overconfidence might be the fact that most of our participants liked the presentation of the spreadsheet. Reithel, Nichols and 
Robinson conducted an experiment \cite{reithel1996experimental} where they asked developers to rate their confidence about the correctness of four spreadsheets. Interestingly, the participants expressed a higher confidence in a large but nicely formatted spreadsheet in contrast to small but poorly formatted spreadsheets. In our study, the majority of the participants liked the presentation of our spreadsheet as well which might have increased their blindness to risks. In this context, it is remarkable that although our control groups rated the maintainability of the spreadsheet as worse than our experiment groups, the control groups were still more confident that their spreadsheet was correct after maintenance. However, such paradox behavior is no exception as Panko has shown in a small meta study \cite{panko2014we}, analyzing other behavioral studies, concluding that humans are generally blind to risks until they occur.

\section{Threats to Validity}

When designing and executing the experiments, we followed the general principles described in \cite{prechelt2001kontrollierte}. Still, the validity of this work might be affected by a number of threats. We split all identified threats into three groups: construct validity (CV), internal validity (IV) and external validity (EV). Below, we provide a brief discussion:

\begin{itemize}
	\item (CV) Spreadsheet Guardian is a conceptual approach and its application requires the presence of concrete testing techniques that follow the approach. For this evaluation we selected one such technique, namely the test scenario technique implemented in SIF. Having collected some evidence that this particular technique basically works one could say that also the conceptual approach works --- at least using this technique. However, drawing general conclusions whether other techniques following Spreadsheet Guardian's concept will work likewise is not possible based on this example alone.

	\item (CV) Spreadsheet Guardian is targeted at typical spreadsheet users. The participants of E3 do not belong to this group. In general, computer science students are believed not to be good at doing work with spreadsheets but are expected to follow a more structured approach than typical spreadsheet users.
	
	One could argue that previous knowledge in unit testing might have benefited computer science students for applying Spreadsheet Guardian - however, our results indicate the opposite.  Overall, it remains questionable if the expected methodological advantage of computer science students can outweigh their spreadsheet skill deficiencies. As the group of typical spreadsheet users is not homogenous either, differences in their background could be influencial as well (i.e. users with a background in engineering might be different from those with a background in social sciences). Therefore, since there is no obvious imbalance between computer science students and typical spreadsheet users, we rate this threat as moderate.
	
	\item (IV) Because most of our participants for E1 and E2 were full time employees, we did the experiments in their homes at varying times of the day. Therefore, we were unable to control factors such as their tiredness, workplace suitability (eight participants wanted to work from their couch or on small kitchen tables) or lighting conditions. We do not regard this threat as critical because environmental factors are not perfect in workplaces either.

	\item (IV) Although the population for E3 had a similar background, using only randomization for shaping the control and experiment groups for E3 may have led to imbalance. While the gender distribution is definitely problematic, other factors like professional work experience and age were comparable. Yet, the differences in confidence ratings are too high to assume their origin in the imbalance between those groups. Therefore, we see this only as a moderate threat.
	
	\item (IV) Many of the participants of E1 and E2 were friends of the first author or friends of the first author's friends. Therefore, one could attest them a friendly attitude towards the experiment. On the other hand, none of the participants of E3 were friends of the first author and some might even have had a hostile attitude towards the experiment as they were forced to participate in at least one experiment (not in ours, but still). Others might have feared that giving negative ratings could be disadvantageous for their relation to the experimentors. Overall, we rate it as a moderate threat that these factors possibly affecting the attitudes of the participants might have impacted their answers to the subjective questions of the experiment. In addition, we do not see clear indications that could confirm this threat in the results either.
	
	\item (EV) We used spreadsheets with seeded defects instead of real spreadsheets. Also, the experiments were tailored to the type of defects our tool is able to detect. On the other hand, this allowed us to make the experiment mostly domain-independent. Also, as the analysis of the defects from the Enron corpus has shown, the defects seeded by us were realistic. Overall, for the aims of our study, we consider this to be an acceptable compromise.
	
	\item (EV) Although we have spent extensive time on recruiting participants for E1 and E2, the sample size remains rather small. This threat, combined with the low number of correctly maintained spreadsheets, keeps us from making statements about Spreadsheet Guardian's effect on maintaining correctness. For E2, we partly mitigated this threat by adding the participants from E3 (but, as discussed, this has brought up new issues).

	\item (EV) While in E1 and E2 we could reject participants, in E3 we had to take every student who registered for our experiment. The higher frustration rate can be clearly seen from our results. However, this also helped us to get a better understanding of sensible requirements for test rule producers. We see this threat to be harmless as long as we cast a skeptical eye on it.
	
	\item (EV) While we believe that SIF's implementation is resistant to structural changes in a spreadsheet to some degree, the issue of test rules becoming outdated due to changes in the logic was not taken care of yet. However, neither SIF's actual resistance to structural changes nor the aspect of logical changes have been touched in the evaluation. However, the same issues arise in traditional software development as well. While we can think of some possibilities for determining cases where users should be alerted about the need to update test specifications, there is no silver bullet for addressing this threat.
	
\end{itemize}

\section{Related Work}

Spreadsheet Guardian is a conceptual approach which is both preventive and detective and employs partly automated inspection techniques -- this is a unique combination. Since we are not aware of other comparable approaches, in the following, we want to compare SIF's test scenario technique with other partly automated detection techniques.

First of all, it is important to know that Spreadsheet Guardian focuses on detecting failures, not faults. Presuming that failures are known, there are several approaches that are able to detect cells which are likely to cause them \cite{abreu2014smelling, hofer2013empirical, abreu2012constraint}. These approaches could be used in conjunction with our approach. 

Among the partly automated inspection techniques, WYSIWYT (\enquote{What you see is what you test}) is probably the most cited one. It was initially developed as a generic testing technique for visual programs \cite{rothermel1998you} and later adapted for spreadsheets \cite{burnett2002testing}. WYSIWYT works as follows: it automatically detects input cells and then populates them with values provided by users or a random generator. Then, users have to state whether they think that values in other cells (i.e. intermediate or result cells) are right or wrong, so WYSIWYT can flag failures and help locating corresponding faults. Spreadsheet Guardian works the other way round and SIF's test scenario technique is a good example of it: it \emph{first} asks the user to compute values outside the spreadsheet and then compares these expectations with values computed by the spreadsheet. We expect that this tackles the overconfidence problem better since humans prefer to make judgments from memory rather than by 
recomputing facts \cite{reder1982plausibility}.

In Ayalew's \enquote{Interval Testing} approach \cite{ayalew2001spreadsheet, ayalew2007user}, users specify ranges of plausible values for input, intermediate and output cells. In a subsequent symbolic execution of the spreadsheet, theoretical violations of these ranges are computed and reported as findings. In contrast to this approach, the test scenario technique implemented in SIF recomputes the spreadsheet in a real and not just a symbolic execution, thus, it is not limited to certain types of supported spreadsheet functions. Another difference is that the test scenario technique detects actual failures and not only theoretical disagreements between the spreadsheet and its test specification. On the other hand, the test scenario technique covers only the few specified test rules while Interval Testing is able to test a much broader range.

The \enquote{EXQUISITE} approach \cite{jannach2014model} is the only approach that implements unit tests for spreadsheets in a comparable fashion to SIF's test scenario technique. However, EXQUISITE relies on automatic detection of cell types and does not allow users to manually specify them \cite{kulesz2014spreadsheet}. Another difference is that EXQUISITE does not execute the inspections continuously in the background. Also, just like WYSIWYT, it does not hide values already present in the spreadsheet at the time of specification which probably also makes it more prone to the overconfidence issue.

While we prefer to employ the test scenario technique as implemented in SIF for the discussed reasons, we also investigated whether the three partially automated testing techniques discussed would be compatible with the conceptual approach proposed by Spreadsheet Guardian:

\begin{itemize}
	
	\item WYSIWYT is not compatible because it does not separate the specification of test rules from the execution of the tests.
	\item Interval Testing is compatible in principle with Spreadsheet Guardian as it separates the specification of test rules from the execution of the tests. It could be integrated in Spreadsheet Guardian provided that the symbolic execution could be done fast enough to provide end-users with live feedback.
	\item EXQUISITE might seem to be compatible with Spreadsheet Guardian at first sight as it implements unit tests in a comparable fashion as SIF's test scenario technique. However, EXQUISITE's tooling turns the spreadsheet environment into a full-blown IDE. This might be adequate for advanced spreadsheet users who want not only fault detection but also debugging support. However, to our knowledge, EXQUISITE is not based on the concept of having separate roles with separate skills for the specification of test rules and the consumption of findings. This makes it incompatible with Spreadsheet Guardian.

\end{itemize}

\section{Lessons Learned}

Before concluding this work, we would like to reflect on our experience with evaluating Spreadsheet Guardian from a methodological perspective. When designing our evaluation, we took into consideration various challenges already reported in the literature. In particular we considered the work by Hofer et al. \cite{hofer2014tool} that is targeted at evaluating approaches for fault-localization in spreadsheets, the work by Panko \cite{panko2014improving} that is targeted at conducting studies about spreadsheet inspections, the work by Powell, Baker and Lawson \cite{powell2008critical} that is targeted at research on spreadsheet errors in general and the work by Prechelt \cite{prechelt2001kontrollierte} that deals with designing controlled experiments in software engineering in general. In the following, we want to reflect on a selection of these and other aspects that we found challenging in particular for our study, how we approached them and what we would recommend to other researchers in this field. To keep 
the reflection lean, we omitted the general challenges of finding appropriate participants and motivating them.

\begin{description}

	\item [Timing and tool implementation] are hard to master. Evaluating a conceptual approach like Spreadsheet Guardian cannot be practically done without having a tool that is representative of the approach. Implementing such a tool can be a long and difficult process (like in our case). The key challenge is to decide when the tool is good enough to use it: In the worst case, you can spend years developing a tool if you want to make it perfect, but if you never do evaluations on a considerable scale you might find out too late that the theory you wanted to evaluate does not work at all. On the other hand, if you build a sloppy tool you might waste a lot of time running meaningless evaluations. In the end, one has to decide when to consider a tool \enquote{good enough} to run the evaluation.
	
	While we had the opportunity to freely decide on the timing for E1 and E2, we were forced to use a certain time slot for E3. During the pilot runs, we found and fixed several severe bugs which could have seriously impacted our study. However, we would argue that you can never have enough pilot runs to find all bugs --- and each pilot you use is one lost participant for your sample. Luckily, we were able to find ad-hoc ways of working around tool issues during the study. Yet, we were frustrated about the bugs since we believe that our evaluation results would have been clearer if our tool had been more mature.

	\item [Using ample sample sizes] comes at the cost of time or quality. Visiting participants at home and observing them individually during the experiments as we did in E1 and E2 gave us researchers more valuable impressions than watching multiple participants doing the task simultenaeously or just analyzing the data. However, having a common time slot where multiple participants show up without all the organizational burden of making individual appointments saves a lot of time and effort. The lesson we learned here is if you want high sample sizes you practically need to make sacrifices in terms of time spent or observation quality. We are sure that we would have lost much direct feedback and the ability to properly react to bugs in the tool implementation if we had conducted the studies as \enquote{mass experiments} e.g. with 30 participants taking the study simultaneously.

	\item [Finding an appropriate difficulty level] needs realistic assumptions about the skills of the participants. The assumptions we had about the spreadsheet skills of the participants in E1 and E2 turned out not to be realistic, as several of them had problems with using the VLOOKUP function. This tought us that even if users have many years of experience with spreadsheets and see themselves as advanced users, a good understanding of maybe less commonly used functions such as VLOOKUP cannot be expected. Regarding the skills of the group which took E3 our estimates were even more off --- we had not expected that computer science students would have such severe problems in using the basic IF function in a spreadsheet. As discussed previously, the difficulty level of our tasks was too high. For future experiments we will try to significantly lower the difficulty because comparing effectiveness is hard if only few participants are able to solve the task correctly at all.
	
	\item [Having an adequate task length] cannot be done without extensive piloting. Initially, we designed our tasks with a duration of 60 minutes in mind. In the pilot runs, this turned out to be rather unrealistic, as the tasks took around 90 minutes. However, most of the participants took between 90 and 120 minutes to finish the experiment. That was more than we wanted, however, we would have needed more pilots to find this out early enough to be able to react. In the end, we think that even 120 minutes were acceptable and adequate in this case. However, we have not figured out a dependable way for making better estimates in the future.
	
	\item [Providing adequate tutorials] can be done with significantly less effort if you avoid providing video tutorials. As we discussed in a separate article \cite{kafer2016best}, there was no significant difference in terms of efficiency and effectiveness of video tutorials over text tutorials in our sample of E3.  However, from our experience, creating and updating text tutorials is much less time-consuming than doing the same for video tutorials. When conducting E1 and E2 (which only had video tutorials), we had to update SIF and, thus, update the tutorials several times as we discovered new issues in each pilot run. The lesson we learned here is that we could have saved a lot of effort on updating tutorials if we had chosen to develop a text tutorial instead of a video tutorial --- without making any heavy sacrifices regarding the effectiveness and efficiency of the tutorial.
	
	\item [Using common benchmark problems] is only an option if other researchers created problems that are aligned with your evaluation goals. Unfortunately, we haven't found any spreadsheet from other experiments that would have been suitable for experimenting with the testing technique implemented in SIF. Yet, we hope that by sharing all (modified and unmodified) spreadsheets from our experiments we can contribute to the selection of available benchmark problems. Of course, we encourage other researchers to use our study objects for their purposes and would be very pleased if other researchers would conduct new experiments based on our study objects.

\end{description}

\section{Conclusion and Future Work}

In this work, we proposed an approach and a theory for supporting the maintenance of spreadsheets in collaborative settings. By continuously applying partly automated inspection techniques that separate the specification of test rules from their execution, Spreadsheet Guardian is able to protect the semantic correctness of spreadsheets to some degree -- at least indirectly.

The results from our study do not indicate that spreadsheets automatically have significantly higher chances of staying correct during maintenance when Spreadsheet Guardian is applied by using SIF's test scenario technique. Yet, our empirical evaluation gives indications that having test rules in a spreadsheet as proposed by Spreadsheet Guardian considerably lowers overconfidence. Hence, it can be assumed that users applying Spreadsheet Guardian would be more careful when taking business-critical decisions based on such spreadsheets. After all, we should not forget that many of the widely reported \enquote{spreadsheet horror stories} did not result from ineffective inspections but simply from overconfidence. Therefore, we recommend to consider Spreadsheet Guardian for any business-critical spreadsheets that are maintained in a collaborative setting. We expect that the importance of supporting such settings will further increase over the next years as businesses adopt new technology that allows end-users to 
work on spreadsheets in the cloud and using mobile devices.

However, as discussed, one could certainly argue that the results obtained from our controlled experiments are not yet convincing. This is why we plan to improve the implementation of the test scenario technique, add more techniques which follow the approach proposed by Spreadsheet Guardian and extend the investigation by running experiments in practical settings as well. Furthermore, it seems worthwhile to conduct a dedicated study on the effects of Spreadsheet Guardian on overconfidence reduction.

%% file: table1.inc
\begin{table}[!htp]

	\centering 
	
	\caption{Theory of the proposed approach (Spreadsheet Guardian)}
	\label{tab:propositionstable}
	
	\normalsize
	
	\begin{tabular}{lp{13cm}}
		\hline 
		\multicolumn{2}{l}{Constructs} \\ 
		\hline
		C1 & Spreadsheet user\\
		C2 & Test rule\\
		C3 & Test rule producer\\
		C4 & Test rule processor\\
		C5 & Findings consumer\\
		C6 & Spreadsheet Guardian\\
		\hline
		\multicolumn{2}{l}{Propositions} \\
		\hline
		P1 & Specifying test rules using Spreadsheet Guardian is feasible for regular spreadsheet users.\\
		P2 & Spreadsheets protected by Spreadsheet Guardian have a higher chance of remaining correct after maintenance.\\
		
		\hline 
		\multicolumn{2}{l}{Explanations} \\ 
		\hline 
		E1 & The specification of test rules is done in the familiar spreadsheet environment. It fits in the spreadsheet paradigm, is easy to learn and does not require computer science knowledge. ($\rightarrow$ P1)\\
		
		E2 & The test rules serve as an implicit specification for (part of) the actual requirements, providing maintainers with more knowledge about the intentions of previous users' development activities. ($\rightarrow$ P2)\\ 
		
		E3 & The clear separation between production and test code makes test code easier to find. This is especially true when compared to test code some advanced spreadsheet users embed directly in their spreadsheets (\cite{hermans2013improving} argues that such practices are common). ($\rightarrow$ P2)\\
		
		E4 & For findings consumers, the reported findings have a high relevance because they are based on test rules specified by humans working with earlier versions of the spreadsheet. ($\rightarrow$ P2)\\ 
		
		
		
		
		\hline 
		\multicolumn{2}{l}{Scope} \\ 
		\hline 
		& Spreadsheets in collaborative environments\\
		\hline
	\end{tabular} 
\end{table}

%% file: table2.inc
\begin{table}[!htp]
	
	\centering 
	
	\caption{Questions from the pre-test relevant for the calculation of the suitability score (The multipliers and the points for each answer were not printed in the original questionnaire.)}
	\label{tab:suitability_score}
	
	\normalsize
	
	\begin{tabular}{p{15cm}}
		
		How long have you been working with Excel? (Multiplier: 5x) \\
		\hline
		- For less than a year (1P)\\
		- For more than year but less than four years (2P)\\
		- For more than four years (3P)\\

\\

		How do you rate your Excel skills? (Multiplier: 5x) \\
		\hline
		- No or little experience (1P)\\
		- Some experience, but still a beginner (2P)\\
		- Plenty of experience, but not an expert (3P)\\
		- Expert (4P)\\

\\
		
		Do you sometimes also create or modify formulas in workbooks? (Multiplier: 20x) \\
		\hline
		- No, if it all I only put in data (1P)\\
		- Yes, I also create or modify formulas (2P)\\
		
\\

		Only if you are interested in the study: Are you currently still in-training or engaged in a job-qualifying study program? Note: If you are working on a Ph.D, please select "no". (Multiplier: 1x) \\
		\hline
		- Yes (1P)\\
		- No (2P)\\	
		
\\

		Only if you are interested in the study: Do you have a working background in computer science of software engineering (e.g. Studies, professional training, perennial practical experience)? (Multiplier: 1x) \\
		\hline
		- Yes (1P)\\
		- No (2P)\\	
		- Partly (please explain briefly) (3P)\\	
			
\\
			
		Do you sometimes use spreadsheets that you did not create yourself (but that were created by someone else)? (Multiplier: 3x) \\
		\hline
		- No, I only use spreadsheets I created myself. (1P)\\
		- Yes, I also use spreadsheets which I did *not* create myself. (2P)\\

\\

		What do you use Excel for? (Multiplier: 3x) \\
		\hline
		- Maintaining data (e.g. lists with names and addresses) (0P)\\
		- Tracking data (e.g. academic records, budgets, ledgers, inventory, ...) (0P)\\	
		- Tracking data (e.g. academic records, budgets, ledgers, inventory, ...) (0P)\\	
		- Data analysis to find out trends and compute forecasts (e.g. finance expenses, trends in numbers of participants, ...)  (1P)\\	
		- Data analysis for decision making and evaluating alternatives (1P)\\
		- Data analysis for scientific analysis (e.g. experiments, measurements, ...) (1P)\\

	\end{tabular} 
\end{table}

%% file: table3.inc
\begin{table}[!htp]
	\centering 
	
	\caption{Ratio between confidence and actual correctness for E2 and E3}
	\label{tab:ratio}
	
	\normalsize
	\begin{tabular}{c|p{1.8cm}|c|c|c|c}
		
		Experiment & Assessment correct? & \# (Very) confident & \# Not confident (at all) & Total & Ratio \\ 
		\hline 
		\multirow{2}{*}{E2e} & \centering Yes & 1 & 3 & 4 & \multirow{2}{*}{1 (4:4)}\\ 
		& \centering No & 4 & 0 & 4 &\\ 
		\hline
		\multirow{2}{*}{E2c} & \centering Yes & 1 & 1 & 2  & \multirow{2}{*}{0.4 (2:5)}\\ 
		& \centering No & 5 & 0 & 5 &\\ 
		\hline
		\multirow{2}{*}{E3e} & \centering Yes & 0 & 10 & 10  & \multirow{2}{*}{2 (10:5)}\\ 
		& \centering No & 5 & 0 & 5 &\\ 
		\hline
		\multirow{2}{*}{E3c} & \centering Yes & 4 & 4 & 8  & \multirow{2}{*}{0.62 (8:13)}\\ 
		& \centering No & 13 & 0 & 13 &\\ 
		
	\end{tabular} 
	
\end{table}

%% file: table4.inc
\begin{table}[!htp]
	
	\centering 
	
	\caption{Extension to the theory of the proposed approach}
	\label{tab:proposition_extend}
	
	\normalsize
	
	\begin{tabular}{lp{13cm}}
		\hline
		\multicolumn{2}{l}{Propositions} \\
		\hline
		P3 & By applying Spreadsheet Guardian, spreadsheet users can reasonably judge the actual correctness of their spreadsheets.\\
		
		\hline 
		\multicolumn{2}{l}{Explanations} \\ 
		\hline 
		E5 & Newly raised findings about violations of test rules specified by spreadsheet users before maintenance are serious indicators for newly introduced faults. ($\rightarrow$P3)\\
		
	\end{tabular} 
\end{table}